\pdfoutput=1 
\documentclass[aps,prd,reprint,superscriptaddress,longbibliography]{revtex4-2}
\usepackage[T1]{fontenc}
\usepackage{braket}
\usepackage{textcomp}
\usepackage{gensymb}
\usepackage{physics}
\usepackage{multirow}
\usepackage{graphicx}% Include figure files
\usepackage{dcolumn}% Align table columns on decimal point
\usepackage{bm}% bold math
\usepackage{amsmath}%[fleqn]
\usepackage{color}
\usepackage{array,url}
\usepackage{xcolor}
\colorlet{BLUE}{blue} % REVTeX uppercases section-title color arguments
\usepackage[caption=false]{subfig}
\usepackage{hyperref}
\hypersetup{
    colorlinks=true,
    linkcolor=black,
    citecolor=black,
    urlcolor=black
}

\def\ber{\begin{eqnarray}}
\def\eer{\end{eqnarray}}
\def\beq{\begin{equation}}
\def\eeq{\end{equation}}

\def\ed{\end{document}}

\makeatletter
\let\jnl@style=\rm
\def\ref@jnl#1{{\jnl@style#1}}

\def\aj{\ref@jnl{AJ}}                   % Astronomical Journal
\def\actaa{\ref@jnl{Acta Astron.}}      % Acta Astronomica
\def\araa{\ref@jnl{ARA\&A}}             % Annual Review of Astron and Astrophys
\def\apj{\ref@jnl{ApJ}}                 % Astrophysical Journal
\def\apjl{\ref@jnl{ApJ}}                % Astrophysical Journal, Letters
\def\apjs{\ref@jnl{ApJS}}               % Astrophysical Journal, Supplement
\def\ao{\ref@jnl{Appl.~Opt.}}           % Applied Optics
\def\apss{\ref@jnl{Ap\&SS}}             % Astrophysics and Space Science
\def\aap{\ref@jnl{A\&A}}                % Astronomy and Astrophysics
\def\aapr{\ref@jnl{A\&A~Rev.}}          % Astronomy and Astrophysics Reviews
\def\aaps{\ref@jnl{A\&AS}}              % Astronomy and Astrophysics, Supplement
\def\azh{\ref@jnl{AZh}}                 % Astronomicheskii Zhurnal
\def\baas{\ref@jnl{BAAS}}               % Bulletin of the AAS
\def\bac{\ref@jnl{Bull. astr. Inst. Czechosl.}}
\def\caa{\ref@jnl{Chinese Astron. Astrophys.}}
\def\cjaa{\ref@jnl{Chinese J. Astron. Astrophys.}}
\def\icarus{\ref@jnl{Icarus}}           % Icarus
\def\jcap{\ref@jnl{J. Cosmology Astropart. Phys.}}
\def\jrasc{\ref@jnl{JRASC}}             % Journal of the RAS of Canada
\def\memras{\ref@jnl{MmRAS}}            % Memoirs of the RAS
\def\mnras{\ref@jnl{MNRAS}}             % Monthly Notices of the RAS
\def\na{\ref@jnl{New A}}                % New Astronomy
\def\nar{\ref@jnl{New A Rev.}}          % New Astronomy Review
\def\pra{\ref@jnl{Phys.~Rev.~A}}        % Physical Review A: General Physics
\def\prb{\ref@jnl{Phys.~Rev.~B}}        % Physical Review B: Solid State
\def\prc{\ref@jnl{Phys.~Rev.~C}}        % Physical Review C
\def\prd{\ref@jnl{Phys.~Rev.~D}}        % Physical Review D
\def\pre{\ref@jnl{Phys.~Rev.~E}}        % Physical Review E
\def\prl{\ref@jnl{Phys.~Rev.~Lett.}}    % Physical Review Letters
\def\pasa{\ref@jnl{PASA}}               % Publications of the Astron. Soc. of Australia
\def\pasp{\ref@jnl{PASP}}               % Publications of the ASP
\def\pasj{\ref@jnl{PASJ}}               % Publications of the ASJ
\def\rmxaa{\ref@jnl{Rev. Mexicana Astron. Astrofis.}}%
\def\qjras{\ref@jnl{QJRAS}}             % Quarterly Journal of the RAS
\def\skytel{\ref@jnl{S\&T}}             % Sky and Telescope
\def\solphys{\ref@jnl{Sol.~Phys.}}      % Solar Physics
\def\sovast{\ref@jnl{Soviet~Ast.}}      % Soviet Astronomy
\def\ssr{\ref@jnl{Space~Sci.~Rev.}}     % Space Science Reviews
\def\zap{\ref@jnl{ZAp}}                 % Zeitschrift fuer Astrophysik
\def\nat{\ref@jnl{Nature}}              % Nature
\def\iaucirc{\ref@jnl{IAU~Circ.}}       % IAU Cirulars
\def\aplett{\ref@jnl{Astrophys.~Lett.}} % Astrophysics Letters
\def\apspr{\ref@jnl{Astrophys.~Space~Phys.~Res.}}
\def\bain{\ref@jnl{Bull.~Astron.~Inst.~Netherlands}}
\def\fcp{\ref@jnl{Fund.~Cosmic~Phys.}}  % Fundamental Cosmic Physics
\def\gca{\ref@jnl{Geochim.~Cosmochim.~Acta}}   % Geochimica Cosmochimica Acta
\def\grl{\ref@jnl{Geophys.~Res.~Lett.}} % Geophysics Research Letters
\def\jcp{\ref@jnl{J.~Chem.~Phys.}}      % Journal of Chemical Physics
\def\jgr{\ref@jnl{J.~Geophys.~Res.}}    % Journal of Geophysics Research
\def\jqsrt{\ref@jnl{J.~Quant.~Spec.~Radiat.~Transf.}}
\def\memsai{\ref@jnl{Mem.~Soc.~Astron.~Italiana}}
\def\nphysa{\ref@jnl{Nucl.~Phys.~A}}   % Nuclear Physics A
\def\physrep{\ref@jnl{Phys.~Rep.}}   % Physics Reports
\def\physscr{\ref@jnl{Phys.~Scr}}   % Physica Scripta
\def\planss{\ref@jnl{Planet.~Space~Sci.}}   % Planetary Space Science
\def\procspie{\ref@jnl{Proc.~SPIE}}   % Proceedings of the SPIE

\begin{document}
\title{A superconducting LC resonator in a multi-tesla magnetic field for tabletop kHz gravitational-wave detection}

\author{D.~Ahn}
\email[Contact author: ]{ahn@pd.infn.it}
\affiliation{INFN, Sezione di Padova, I-35131 Padova, Italy}

\author{C.~Braggio}
\affiliation{Dipartimento di Fisica e Astronomia ``Galileo Galilei'', {Universit\`a} degli Studi di Padova, I-35131 Padova, Italy}
\affiliation{INFN, Sezione di Padova, I-35131 Padova, Italy}

\author{G.~Carugno}
\affiliation{INFN, Sezione di Padova, I-35131 Padova, Italy}

\author{D.~D'Agostino}
\affiliation{INFN, Gruppo Collegato di Salerno, Sezione di Napoli, I-84084 Fisciano, Salerno, Italy}

\author{R.~Di Vora}
\affiliation{INFN, Laboratori Nazionali di Legnaro, I-35020 Legnaro, Padova, Italy}

\author{P.~Falferi}
\affiliation{Istituto di Fotonica e Nanotecnologie - CNR and Fondazione Bruno Kessler, I-38123 Povo, Trento, Italy}
\affiliation{INFN, TIFPA, I-38123 Povo, Trento, Italy}

\author{A.~Ortolan}
\affiliation{INFN, Laboratori Nazionali di Legnaro, I-35020 Legnaro, Padova, Italy}

\author{G.~Ruoso}
\affiliation{INFN, Laboratori Nazionali di Legnaro, I-35020 Legnaro, Padova, Italy}

\author{M.~L.~Ruggiero}
\affiliation{Dipartimento di Matematica ``Giuseppe Peano'', {Universit\`a} degli Studi di Torino, I-10123 Torino, Italy}
\affiliation{INFN, Sezione di Torino, I-10125 Torino, Italy}

\date{\today}% It is always \today, today,
%              %  but any date may be explicitly specified

\begin{abstract}
We present {a concept for a superconducting inductance--capacitance (LC) resonator} immersed in a magnetic field for gravitational-wave detection in the kHz band. In the long-wavelength limit, a gravitational wave in a static magnetic field induces an oscillating magnetic-field perturbation proportional to the {static magnetic field applied to the resonator} and to the gravitational-wave strain. We show that a superconducting pickup circuit can convert this perturbation into a flux signal read out by {a superconducting quantum interference device (SQUID)-based or quantum-limited amplifier chain}. Inspired by resonant-mass detectors, we develop a {three-loop coupled-circuit model} and derive the equivalent strain-noise spectrum, including thermal noise, SQUID backaction, and measurement imprecision. Prototype measurements demonstrate a quality factor of $Q\simeq5\times10^5$, while $Q=10^7$ is considered as a future development target. Under the assumed flux coupling and an idealized quantum-limited SQUID noise model, optimized circuits yield projected strain sensitivities of order $10^{-21}~{\rm Hz}^{-1/2}$ over 100~Hz bands in the 1–20 kHz range. Increasing the active high-field volume and improving quantum-limited readout could enable further sensitivity gains. This compact, tunable detector concept could complement interferometric gravitational-wave searches and motivate investigations of high-frequency astrophysical sources.
\end{abstract}

\maketitle

\section{Introduction}
\label{sec_introduction}

The direct observation of gravitational waves has opened a new observational window on the Universe and has provided a direct experimental test of dynamical strong-field gravity~\cite{Abbott2016GRTest}.
The first detection, GW150914, established the existence of stellar-mass binary-black-hole mergers and demonstrated that interferometric detectors can measure spacetime strain at astrophysical distances~\cite{Abbott2016blz}.
The subsequent observation of GW170817 from a binary-neutron-star inspiral, together with electromagnetic counterparts, inaugurated multi-messenger gravitational-wave astronomy and connected compact-binary dynamics to nuclear physics, relativistic astrophysics, gamma-ray bursts, and cosmology~\cite{Abbott2017bns,Abbott2017MultiMessenger,Abbott2018EOS,Abbott2017H0}.
These discoveries motivate extending gravitational-wave detection not only in sensitivity, but also in frequency coverage and detector principle.
Different frequency bands probe different source classes and different noise mechanisms, so compact resonant detectors can provide a useful complement to kilometer-scale laser interferometers~\cite{Sathyaprakash2009LRR,Aguiar2010zz}.

The kilohertz band is particularly important because it contains information that is difficult to access from the inspiral signal alone~\cite{Rezzolla2016Spectral}. In binary-neutron-star mergers, the post-merger remnant can emit gravitational waves at frequencies of order a few kilohertz, carrying information on the finite-temperature equation of state, the remnant lifetime, possible phase transitions in dense matter, and the threshold between hypermassive-neutron-star formation and prompt collapse~\cite{Hotokezaka2011EOS,Fields2023Thermal,Bauswein2019PhaseTransition,Bauswein2013PromptCollapse,Koeppel2019Threshold}.
More generally, the 2--20~kHz range has been discussed as a target band for compact resonant detectors, including possible signals from binary-neutron-star merger and post-merger remnants, subsolar-mass primordial-black-hole mergers, collapsing stellar cores, and nearly monochromatic emission associated with ultralight boson clouds around rotating black holes~\cite{Naroznik2023bqa}.
Large-scale interferometers are optimized near the tens-to-hundreds of hertz band, while their strain sensitivity degrades at high frequencies where quantum shot noise becomes increasingly important~\cite{Aasi2015lqx,Buikema2020O3}.
This motivates detector concepts optimized for narrow-band or tunable searches in the kHz region, where a compact resonant system can trade broadband coverage for enhanced response over a targeted frequency interval.

Resonant-mass detectors provide an important historical and technical foundation for such efforts. Cryogenic bar detectors formed an international network for burst gravitational-wave searches before the first interferometric detections, while also establishing key techniques for precision, low-noise measurements~\cite{Aguiar2010zz,Astone2010zz}. Among them, AURIGA, the ultracryogenic resonant-bar detector operated at INFN Laboratori Nazionali di Legnaro, represented a state-of-the-art implementation, combining a massive acoustic antenna with a capacitive transducer and SQUID readout. One of its major achievements was a three-mode readout architecture, in which the mechanical modes of the bar and transducer were matched to a high-quality electrical resonator and a two-stage SQUID amplifier. This configuration widened the useful detection bandwidth and achieved a one-sided strain noise below $10^{-20}\,\mathrm{Hz}^{-1/2}$ over approximately 110~Hz around the bar resonance~\cite{Baggio2005qw}. The same experimental platform also demonstrated active feedback cooling of macroscopic electromechanical normal modes through the SQUID input circuit, reducing the effective mode temperature from a 4.2~K bath to 0.17~mK~\cite{Vinante2008zz}. Its later use in searches for ultralight scalar dark matter further showed that resonant detectors can be repurposed as precision probes of physics beyond the Standard Model~\cite{Branca2016rez}.

The present work revisits this resonant-detector tradition from an electromagnetic-circuit perspective. In a complementary high-field configuration, a gravitational wave propagating through a static magnetic field induces an oscillating magnetic-field perturbation proportional to the external field and to the strain, which a superconducting pickup loop can convert into a resonant flux signal in an LC circuit~\cite{Fortini1996, Domcke2023Selection, Domcke2025}.
Compared with a massive acoustic bar, such an electromagnetic resonator can be made compact, tunable, and naturally compatible with millikelvin cryogenics, high magnetic fields, and quantum-limited readout~\cite{Clerk2010rmp,Kolevatov2026HighQLumped}. This route connects the legacy of resonant-mass gravitational-wave detection with superconducting lumped-element techniques developed for dark-matter searches. Existing implementations establish the relevant circuit ingredients: static-field pickup, superconducting lumped-element resonance, shielding and cryogenic operation, and SQUID-based flux readout~\cite{Kahn2016ABRACADABRA,Brouwer2022DMRadiom3,Pappas2025ABRACADABRAGW}. We therefore use benchmark conditions compatible with these circuit technologies, including operation around 10~mK, magnets in the 9--14~T range {with a bore size of around 10~cm}, superconducting resonators, and low-noise flux readout, while applying them to a gravitational-wave-induced flux drive rather than to an axion or hidden-photon source. Prototype superconducting LC resonators have already been characterized in the kHz range, with measured quality factors reaching $Q\simeq5\times10^{5}$ at 1.5~kHz. Higher-$Q$ superconducting or high-temperature-superconducting implementations, optimized coupling, and feedback-assisted mode cooling then provide possible routes toward a narrow-band, tunable, tabletop detector targeting $10^{-21}\,\mathrm{Hz}^{-1/2}$ strain sensitivity in the kHz band.

\section{Detection principle}

We consider a superconducting solenoidal pickup circuit placed in a static magnetic field $\mathbf B_0=B_0\hat{\mathbf z}$ inside a region much smaller than the gravitational wavelength. The detector frame is denoted by $\mathcal D=(\hat{\mathbf x},\hat{\mathbf y},\hat{\mathbf z})$, where $\hat{\mathbf z}$ is parallel to the solenoid axis and the pickup loop is oriented to measure magnetic flux through the corresponding area. The gravitational wave is described in a wave frame $\mathcal G=(\hat{\mathbf X},\hat{\mathbf Y},\hat{\mathbf Z}$), with the direction of propagation $\hat{\mathbf n} \parallel \hat{\mathbf Z}$. The relative orientation of the two frames is specified by the Euler angles $(\theta,\phi,\psi)$, where
\begin{equation}
\begin{split}
\hat{\mathbf n}
=
(&\sin\theta\cos\phi,
\sin\theta\sin\phi,
\cos\theta)_{\mathcal D}.
\end{split}
\label{eq_gw_direction}
\end{equation}
Here $\psi$ is the polarization angle in the plane orthogonal to $\hat{\mathbf n}$. A convenient basis in this orthogonal plane is
\begin{align}
\hat{\mathbf e}_\theta
&=
(
\cos\theta\cos\phi,
\cos\theta\sin\phi,
-\sin\theta),
\\
\hat{\mathbf e}_\phi
&=
(-\sin\phi,
\cos\phi,
0).
\end{align}
The polarization basis is then defined by
\begin{align}
\hat{\mathbf X}
&=
\cos\psi \hat{\mathbf e}_\theta
+
\sin\psi \hat{\mathbf e}_\phi,
\\
\hat{\mathbf Y}
&=
-\sin\psi \hat{\mathbf e}_\theta
+
\cos\psi\hat {\mathbf e}_\phi .
\end{align}
The transverse-traceless polarization tensors are 
\begin{align}
e^+_{ij}
=
\hat X_i\hat X_j-\hat Y_i\hat Y_j,
e^\times_{ij}
=
\hat X_i\hat Y_j+\hat Y_i\hat X_j,
\end{align}
and the metric perturbation is written as
\begin{equation}
h_{ij}(t)
=
h_+(t)e^+_{ij}
+
h_\times(t)e^\times_{ij} \ \ .
\end{equation}

{In the long-wavelength limit} the metric perturbation is nearly uniform over the detector volume. The detector response may be conveniently described in the local proper detector frame. This is naturally represented by {Fermi normal coordinates} attached to the detector world-line, in which the origin is locally inertial, time is measured by a {clock at the origin} (detector proper time),  and the measured electromagnetic field is obtained by projecting $F_{\mu\nu}$ onto the detector tetrad~\cite{Rakhmanov2014}. Although components of $F_{\mu\nu}$ 
%$\mathbf E$ and $\mathbf B$ 
depend on the coordinate system and tetrad, the magnetic flux through a physical pickup loop $\Phi_B$ is a scalar, gauge-independent quantity as its perturbation $\delta \Phi_{\rm GW}$ turns out to be proportional to the geometric curvature of spacetime~\cite{Ruggiero_2020,Park_2022_EM_invariants_GW}. In differential-form notation it is 
\begin{equation}
\Phi_B(t)
=
\int_{\Sigma(t)} F ,
\label{eq_flux_two_form}
\end{equation}
where $F=dA$ is the electromagnetic field two-form and $\Sigma(t)$ is the physical surface bounded by the superconducting loop at detector proper time $t$. This expression is invariant under electromagnetic gauge transformations $A\rightarrow A+d\Lambda$, since $F\rightarrow F+d^2\Lambda=F$. It is also invariant under spacetime gauge transformations (i.e. infinitesimal coordinate transformations), provided that the field and the physical loop surface are transformed together
\begin{equation}
\int_{\Sigma}F
=
\int_{\varphi^{-1}(\Sigma)}\varphi^*F .
\label{eq_flux_diffeomorphism}
\end{equation}
Thus the gravitational-wave-induced flux is an observable independent of the coordinate system in which it is computed. ~\cite{StewartWalker1974,BruniSonego1999,Park_2022_EM_invariants_GW}. A calculation that consistently accounts for field configuration and physical loop surface yields the same flux perturbation for any infinitesimal coordinate transformation.
%In practice the two descriptions distribute the same physical effect differently: in TT coordinates part of the response appears through the metric and the proper area of the loop, whereas in Fermi-normal coordinates the apparatus geometry is locally rigid and the response appears through the locally measured electromagnetic field.
The gravitational-wave-induced flux at the first order in $h_{+,\times}$ can therefore be written directly in terms of the scalar detector response
\begin{equation}
\begin{split}
\delta \Phi_h (t)
&=
N\int_{A_{\rm p}}
\delta\mathbf B(t)\cdot d\mathbf A
\\
&\simeq
N A_{\rm p} B_0
\big[
F_+ \ h_+(t) + 
F_\times \ h_\times(t)
\big] \ , \\ 
\end{split}
\label{eqfluxgw}
\end{equation}
where $N$ is the number of turns, $A_{\rm p}$ is the effective pickup area, and  ${F_{+,\times}} $ is the antenna pattern defined as the contraction of the detector tensor with the gravitational-wave polarization tensor.

\begin{figure}[t]
\centering
\includegraphics[width=\linewidth]{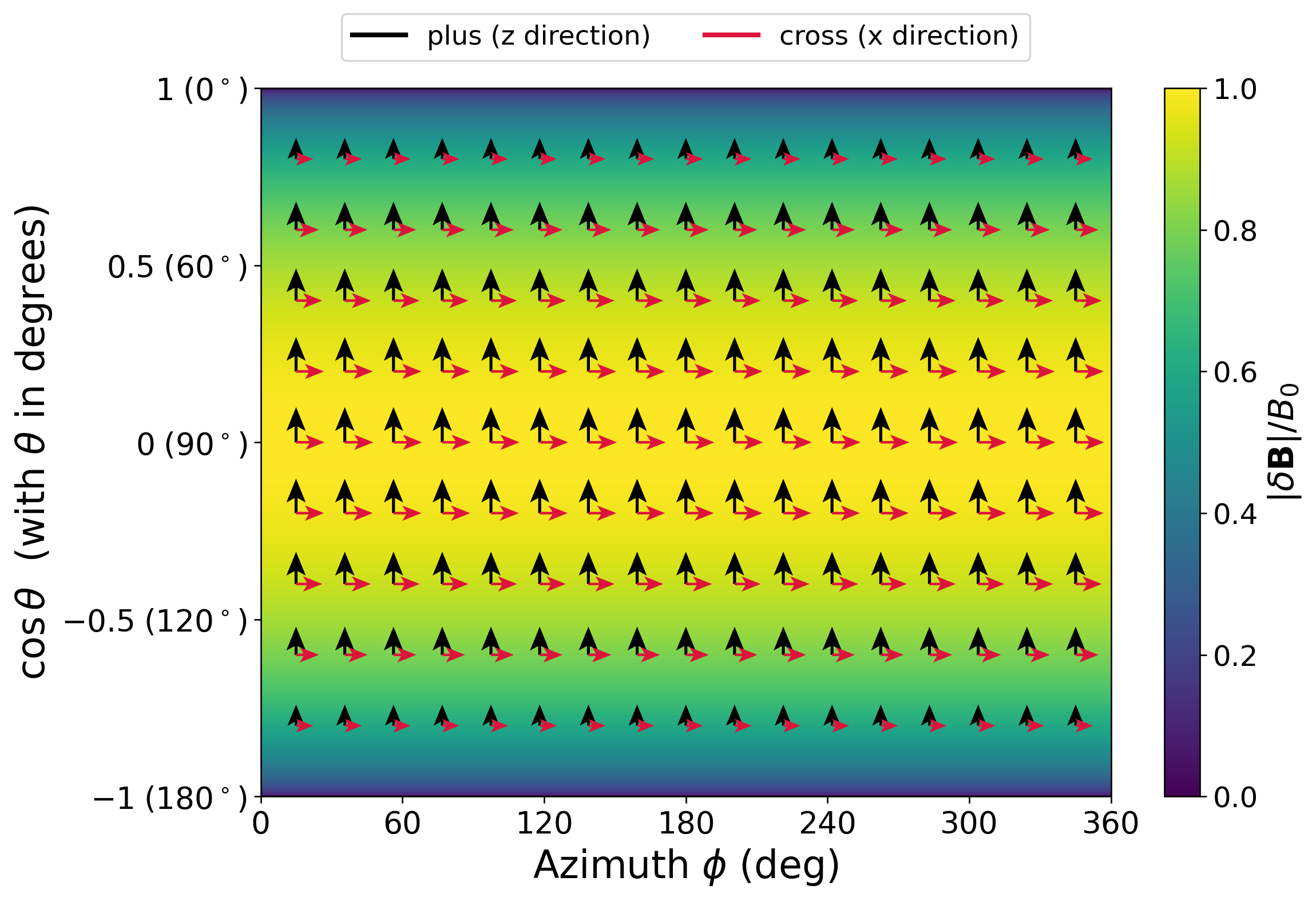}
\caption{
Antenna pattern of the magnetic field response to the two gravitational-wave polarizations. The black upward and red rightward arrows represent the induced magnetic field directions for the $+$ and {$\times$} polarizations, respectively. Note that they are orthogonal as expected.
}
\label{fig:antenna_pattern}
\end{figure}
% V9 addition: mechanical-response assumption and cancellation qualification.
%{ The magnet and pickup loop are assumed to be mechanically distinct bodies, with the relevant mechanical resonances outside the detector bandwidth. In this configuration, due to  gravitational wave interaction, an oscillating magnetic field is superimposed to the static field and collected resonantly by the pickup loop. The high-Q factor of the pickup circuit ensures that the oscillating magnetic filed is the dominant contribution of the detector response to a gravitational wave.   ~\cite{Domcke2023Selection,Domcke2024Weber}.
%{ For mechanically distinct magnet and pickup bodies, Equation~\eqref{eqfluxgw} can be viewed as a flux generator associated with the incoming gravitational wave driving the pickup circuit.}

For the axisymmetric solenoidal geometry considered here, the detector tensor and antenna pattern read 
\begin{equation}
D^{ij}=\hat z^i\hat z^j \ ,
\end{equation}
\begin{equation}
F_{+,\times }(\theta,\phi,\psi)
=
D^{ij}e^{+,\times }_{ij} \ , 
\label{eq_detector_tensor_response}
\end{equation}
where $\hat z^i$ is the unit vector parallel to the pickup solenoid. 
%and $\theta,\phi,\psi$ are the Euler angles relating the detector and Transverse-Traceless (TT) reference systems.  
%Because the perturbation $\delta \Phi_{\rm GW}(t)$ is a scalar contraction of geometrically defined tensors, it is ... provided that the long wavelength approximation holds. 
Once the detector axis and the wave direction are physically specified, the coordinate components of $h_{ij}$, $\delta\mathbf B$, or the loop area $\Sigma$ depend on reference systems, but the contracted response entering Eq.~\eqref{eqfluxgw} does not.

Using the basis above, one obtains 
\begin{align}
F_+(\theta,\phi,\psi)
&=
\hat z^i\hat z^j e^+_{ij}
=
\sin^2\theta\cos 2\psi,
\label{eq_Fplus_scalar}
\\
F_\times(\theta,\phi,\psi)
&=
\hat z^i\hat z^j e^\times_{ij}
=
-\sin^2\theta\sin 2\psi \ ,
\label{eq_Fcross_scalar}
\end{align}
{where $\theta$ and $\phi$ specify the gravitational-wave propagation direction; the corresponding source direction on the sky is opposite to $\hat{\mathbf n}$.} Figure \ref{fig:antenna_pattern} shows an antenna pattern of the detector. The absence of explicit $\phi$ dependence is a consequence of the azimuthal symmetry of an ideal solenoidal detector around $\hat{\mathbf z}$. A non-axisymmetric pickup geometry would replace $D^{ij}=\hat z^i\hat z^j$ by the appropriate detector tensor and would in general restore explicit dependence on $\phi$. Equations~\eqref{eq_Fplus_scalar} and~\eqref{eq_Fcross_scalar} show that the detector is blind to waves propagating along the solenoid axis and has maximum response for waves incident perpendicular to the static magnetic field. Equivalently, the response is controlled by the transverse part of the static field relative to the gravitational-wave propagation direction. 

The leading gravitational perturbation of the locally measured magnetic field is linear in both $B_0$ and the strain. In the polarization convention above,
\begin{equation}
\begin{split}
\frac{\delta B_z}{B_0}
=
&h_+(t)\sin^2\theta\cos 2\psi
\\
&-
h_\times(t)\sin^2\theta\sin 2\psi
,
\end{split}
\label{eq_dBz_response}
\end{equation}
which is the near-field representation of the flux response in Eq.~\eqref{eqfluxgw}. The conversion of gravitational radiation into electromagnetic perturbations in a static magnetic field is closely related to the inverse Gertsenshtein effect \cite{Marklund1999,Domcke2025}. The relevant observable is the resulting near-field magnetic flux through the pickup loop, rather than outgoing electromagnetic radiation. Here, the magnet and pickup loop are assumed to be mechanically distinct bodies, with their relevant mechanical resonances lying outside the detection bandwidth. In this configuration, the interaction with a gravitational wave induces an oscillating magnetic field that is superimposed on the static field and resonantly detected by the pickup loop. The high quality factor of the pickup circuit enhances the response to the oscillating magnetic field, making it the dominant contribution to the detector response to a gravitational wave~\cite{Domcke2023Selection,Domcke2024Weber}.

The induced flux provides an electromotive drive to the superconducting LC circuit. Faraday induction gives the equivalent source voltage in the frequency domain,
\begin{equation}
V_h (\omega)
=
-i\omega \Phi_h (\omega) .
\label{eqsourcevoltage}
\end{equation}
For a series RLC circuit, the induced current is
\begin{equation}
\begin{split}
I_h (\omega) 
=
\frac{V_h (\omega)}
{R+i\omega L+1/(i\omega C)} .
\end{split}
\label{eqcurrentresponse}
\end{equation}
Near the resonance $\omega_0=1/\sqrt{LC}$, the reactive impedance is minimized and the high quality factor enhances the current response to the applied flux.

Gravitational waves can also modulate the circuit geometry and hence its inductance and capacitance \cite{Fortini1996}. The sensitivity estimates in this work retain the external flux drive in Eqs.~\eqref{eqfluxgw} and \eqref{eqsourcevoltage} as the assumed signal channel, while possible parametric variations of the circuit parameters are not explicitly included in the benchmark calculation. Representing the source in the inductive flux equation is an equivalent-circuit description of the external drive, rather than an assumption that gravitational waves act only on the inductor. The resulting circuit response is transferred to a SQUID-based or quantum-limited readout.

This architecture combines resonant enhancement, electrical tunability, and superconducting readout in a compact detector whose dimensions are primarily determined by the magnet and shielding geometry. Adjusting the inductance or capacitance allows narrow-band searches at different frequencies in the kHz range. The approach is complementary to cavity-based electromagnetic gravitational-wave detector proposals \cite{Herman2023}, while employing a high-field superconducting LC resonator and low-noise flux readout.

\begin{figure*}[t]
\centering
\includegraphics[width=0.95\linewidth]{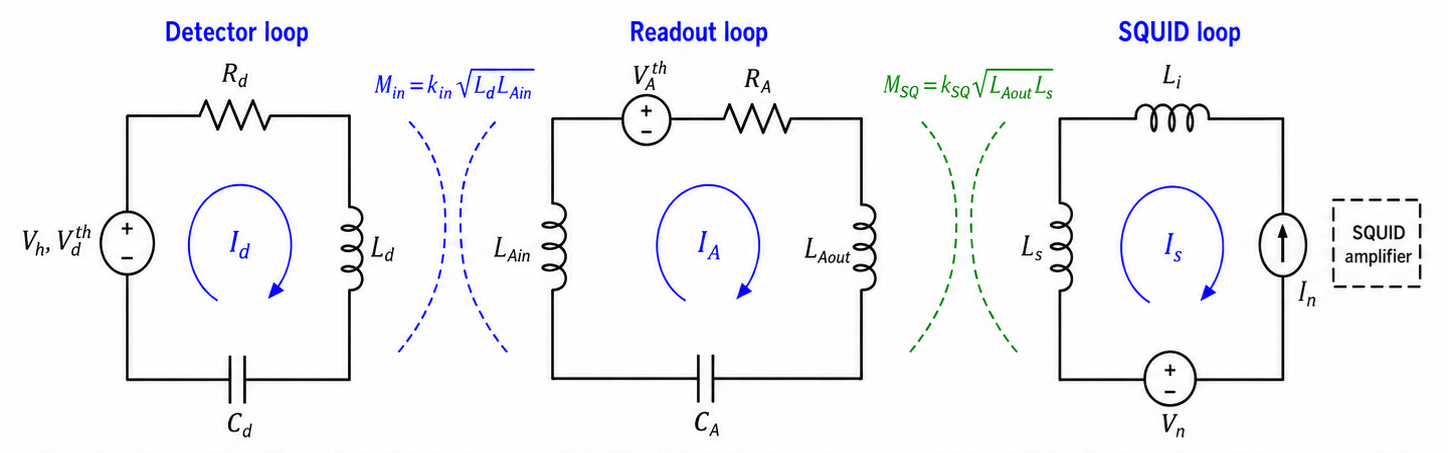}
\caption{
Three-loop direct-coupled detector, readout, and SQUID circuit. The detector inductor is coupled directly to the readout input inductor, and the readout output inductor is coupled to the SQUID loop.
}
\label{fig_three_loop_direct_coupling}
\end{figure*}

\section{LC response and readout}
\label{sec_readout}

Figure~\ref{fig_three_loop_direct_coupling} shows the three-loop circuit model used to describe the LC response and the SQUID readout. The detector loop consists of a superconducting pickup coil with inductance $L_d$, capacitance $C_d$, and residual series resistance $R_d$. The readout loop consists of the input and output inductors $L_{A\mathrm{in}}$ and $L_{A\mathrm{out}}$, the tuning capacitor $C_A$, and the loss resistance $R_A$. The SQUID loop is modeled by the SQUID-loop inductance $L_s$ and the input inductance $L_i$. The detector loop is coupled directly to the readout input inductor through $M_{\rm in}$, while the readout output inductor is coupled to the SQUID loop through $M_{\rm SQ}$. This structure is the electromagnetic analogue of the coupled resonant readout used in AURIGA, where the transducer electrical mode and the SQUID amplifier were tuned to widen the useful detection bandwidth of a resonant detector~\cite{Baggio2005qw,Vinante2002cqg}. The use of a SQUID as a low-noise current and flux amplifier follows the standard treatment of SQUID readout and quantum-limited linear amplification~\cite{ClarkeBraginski2004,Clerk2010rmp,Falferi1998apl,Kolevatov2026SQUIDReadout}.

The loop-current vector is written as
\begin{equation}
\bm I=(I_d,I_A,I_s)^T.
\end{equation}

{
{The gravitational-wave signal is represented as an effective external-flux source in the detector loop. Here $A_{\rm eff}\equiv A_{\rm p}$ and $B\equiv B_0$, and the scalar notation is defined by $F(\theta,\psi)h\equiv F_+h_+ + F_\times h_\times$ for the gravitational-wave polarization components considered above.} For a pickup coil with $N$ turns, effective area $A_{\rm eff}$, static magnetic field $B$, and antenna factor $F(\theta,\psi)$, the signal flux is
\begin{equation}
\Phi_h(\omega)
=
N B A_{\rm eff}F(\theta,\psi)h(\omega)
\equiv
\Lambda h(\omega).
\label{eq_signal_flux_readout}
\end{equation}
where $\Lambda\equiv N B A_{\rm eff}F(\theta,\psi)$. A residual magnetic-field fluctuation $\delta B(\omega)$ would produce an additional background flux
\begin{equation}
\Phi_{\rm bg}(\omega)
=
N A_{\rm eff}\,\delta B(\omega),
\label{eq_background_flux_readout}
\end{equation}
which is not included in the ideal strain-transfer function derived below and is discussed separately in Sec.~\ref{sec_magnet_requirement}. Using the Fourier convention adopted in Eq.~\eqref{eqsourcevoltage}, the corresponding gravitational-wave source voltage is
\begin{equation}
V_h(\omega)
=
-i\omega\Phi_h(\omega)
=
-i\omega\Lambda h(\omega).
\label{eq_Vh_three_loop}
\end{equation}
The overall phase of the transfer function depends on the Fourier convention, whereas all power spectral densities and the magnitude $|H_{\Phi h}|^2$ are convention independent.
}

This source term should be understood as a flux drive produced by the gravitational-wave-induced magnetic perturbation through the pickup coil. A gravitational wave can in principle modulate the circuit parameters $R$, $L$, and $C$ in the long-wavelength limit~\cite{Fortini1996,Montanari1996}. In the present high-field pickup configuration, however, the leading signal channel is the external flux through the detector coil. It is therefore convenient to place the signal source in the detector-loop flux equation. This does not mean that only the inductor carries a physical voltage. The same loop current flows through $R_d$, $L_d$, and $C_d$, while the gravitational-wave-induced magnetic flux appears as the driving electromotive force in the inductive pickup loop. Parametric modulations of $R_d$, $L_d$, and $C_d$ can be included as higher-order corrections to the same circuit model.

{
The detector resonance is fixed at the target angular frequency
\begin{equation}
\omega_0=2\pi f_0,
\qquad
C_d=\frac{1}{\omega_0^2L_d},
\qquad
R_d=\frac{\omega_0L_d}{Q_d}.
\label{eq_detector_resonance_parameters}
\end{equation}
The readout resonance is allowed to be slightly detuned from the detector resonance,
\begin{equation}
\omega_A
=
2\pi\left(f_0+\Delta f_A\right),
\label{eq_readout_detuned_frequency}
\end{equation}
where $\Delta f_A$ is treated as a circuit-design parameter. Defining
\begin{equation}
L_A\equiv L_{A\mathrm{in}}+L_{A\mathrm{out}},
\end{equation}
the readout capacitance and resistance are
\begin{equation}
C_A=\frac{1}{\omega_A^2L_A},
\qquad
R_A=\frac{\omega_A L_A}{Q_A}.
\label{eq_readout_resonance_parameters}
\end{equation}
The special case $\Delta f_A=0$ recovers exact tuning of the detector and readout resonances. Allowing a small nonzero detuning provides an additional degree of freedom for balancing the two coupled-mode sensitivity minima over the selected analysis band.
}

The loop impedances are
\begin{align}
Z_d(\omega)
&=
R_d+i\omega L_d+\frac{1}{i\omega C_d},
\label{eq_Zd_three_loop}
\\
Z_A(\omega)
&=
R_A+i\omega L_A+\frac{1}{i\omega C_A},
\label{eq_ZA_three_loop}
\\
Z_s(\omega)
&=
i\omega(L_s+L_i).
\label{eq_Zs_three_loop}
\end{align}

The mutual inductances are
\begin{equation}
M_{\rm in}
=
k_{\rm in}\sqrt{L_dL_{A\mathrm{in}}},
\qquad
M_{\rm SQ}
=
k_{\rm SQ}\sqrt{L_{A\mathrm{out}}L_s}.
\label{eq_mutual_inductances_three_loop}
\end{equation}

The three-loop equations are then
\begin{equation}
\begin{pmatrix}
Z_d&i\omega M_{\rm in}&0\\
i\omega M_{\rm in}&Z_A&i\omega M_{\rm SQ}\\
0&i\omega M_{\rm SQ}&Z_s
\end{pmatrix}
\begin{pmatrix}
I_d\\
I_A\\
I_s
\end{pmatrix}
=
\begin{pmatrix}
V_h\\
0\\
0
\end{pmatrix}.
\label{eq_three_loop_matrix}
\end{equation}

For compact notation we define
\begin{align}
{\cal A}_2(\omega)
&=
Z_dZ_A+\omega^2M_{\rm in}^2,
\label{eq_A2_three_loop}
\\
{\cal D}_3(\omega)
&=
Z_s{\cal A}_2+\omega^2M_{\rm SQ}^2Z_d.
\label{eq_D3_three_loop}
\end{align}

The measured SQUID flux is taken to be
\begin{equation}
\Phi_{\rm meas}=L_i I_s.
\label{eq_measured_squid_flux}
\end{equation}

{
Solving Eq.~\eqref{eq_three_loop_matrix} gives
\begin{equation}
I_s(\omega)
=
-\frac{\omega^2M_{\rm in}M_{\rm SQ}}{{\cal D}_3(\omega)}
V_h(\omega).
\label{eq_Is_three_loop}
\end{equation}
Consequently, apart from a Fourier-convention-dependent overall phase,
\begin{equation}
\frac{\Phi_{\rm meas}(\omega)}{h(\omega)}
\propto
\frac{\omega^3L_i\Lambda M_{\rm in}M_{\rm SQ}}
{{\cal D}_3(\omega)},
\end{equation}
and the strain-to-flux power transfer function is
\begin{equation}
|H_{\Phi h}(\omega)|^2
=
\left|
\frac{\Phi_{\rm meas}}{h}
\right|^2
=
\frac{
L_i^2\omega^6\Lambda^2
M_{\rm in}^2M_{\rm SQ}^2
}{
|{\cal D}_3(\omega)|^2
}.
\label{eq_Hphih_three_loop}
\end{equation}
The factor $\omega^6$ follows from the two mutual-inductive transfers, which contribute $\omega^4$, and from the flux-drive voltage $V_h=-i\omega\Lambda h$, which contributes one additional factor $\omega^2$ to $|H_{\Phi h}|^2$.
}

The poles of ${\cal D}_3$ correspond to the {normal-mode poles of the coupled detector--readout--SQUID circuit}. By tuning the coupling strengths, the readout parameters, and the small readout detuning $\Delta f_A$, the response can be split into nearby coupled-mode features around the target frequency. This is useful because a single high-$Q$ resonance is converted into a wider useful detection band, in analogy with the multi-mode bandwidth-widening strategy demonstrated in AURIGA~\cite{Baggio2005qw}.

The output noise measured by the SQUID is expressed as {a flux-noise power spectral density (PSD) $S_{\Phi\Phi}^{\rm out}$}. For an equivalent gravitational-wave strain-noise spectrum $S_{hh}^{\rm total}$, the corresponding output flux noise is
\begin{equation}
S_{\Phi\Phi}^{\rm out}(\omega)
=
|H_{\Phi h}(\omega)|^2
S_{hh}^{\rm total}(\omega).
\label{eq_output_flux_noise}
\end{equation}

It is important to distinguish the output flux noise from the input-referred strain noise. The output spectrum $S_{\Phi\Phi}^{\rm out}$ can show resonant peaks because the coupled-circuit transfer function enhances both signal and output noise near the normal-mode poles. The physically relevant strain sensitivity is obtained by dividing by the same transfer function,
\begin{equation}
S_{hh}^{\rm total}(\omega)
=
\frac{
S_{\Phi\Phi}^{\rm out}(\omega)
}{
|H_{\Phi h}(\omega)|^2
}.
\label{eq_input_referred_noise}
\end{equation}
Thus the common pole structure of the signal response and output-noise spectrum is removed in the input-referred spectrum. Residual frequency dependence in $S_{hh}^{\rm total}$ comes from the distinct detector thermal, readout thermal, backaction, and additive contributions after they are referred to equivalent strain.

{
We separate the detector thermal contribution from the noise added by the readout and SQUID chain:
\begin{align}
S_{hh}^{\rm added}
&=
S_A+S_{\rm ba}+S_{\rm add},
\label{eq_added_shh}
\\
S_{hh}^{\rm total}
&=
S_d+S_{hh}^{\rm added}
\nonumber\\
&=
S_d+S_A+S_{\rm ba}+S_{\rm add}.
\label{eq_total_shh}
\end{align}
This distinction is useful because the detector thermal floor is fixed by the detector geometry, temperature, and quality factor, whereas the coupled-circuit optimization acts primarily on the added-noise spectrum.
}

The detector thermal contribution is
\begin{equation}
S_d
=
\frac{4k_BT_dR_d}{\omega^2\Lambda^2},
\label{eq_detector_thermal}
\end{equation}
and the readout-resonator thermal contribution is
\begin{equation}
S_A
=
\frac{
|Z_d|^2\,4k_BT_AR_A
}{
\omega^4\Lambda^2M_{\rm in}^2
}.
\label{eq_readout_thermal}
\end{equation}
The effective SQUID readout model represents voltage backaction as a series voltage source in the SQUID-side loop, while additive current noise is referred directly to the measured SQUID-side current. These noise sources are assumed to be uncorrelated and are referred to equivalent strain using the same signal transfer function. The SQUID backaction and additive (measurement-imprecision) noise terms are modeled by equivalent voltage- and current-noise spectral densities $S_{vv}$ and $S_{ii}$, following the linear-amplifier description of SQUID noise~\cite{Falferi1998apl,Clerk2010rmp,Kolevatov2026SQUIDReadout}:
\begin{align}
S_{\rm ba}
&=
\frac{
|{\cal A}_2|^2S_{vv}
}{
\omega^6\Lambda^2
M_{\rm in}^2M_{\rm SQ}^2
},
\label{eq_squid_backaction_noise}
\\
S_{\rm add}
&=
\frac{
|{\cal D}_3|^2S_{ii}
}{
\omega^6\Lambda^2
M_{\rm in}^2M_{\rm SQ}^2
}.
\label{eq_squid_imprecision_noise}
\end{align}
Here, ``additive'' refers specifically to the SQUID current-noise contribution associated with $S_{ii}$. The same contribution is commonly termed measurement imprecision in the general linear-detector formalism, whereas $S_{hh}^{\rm added}$ denotes the total readout-added strain noise, $S_A+S_{\rm ba}+S_{\rm add}$, used in the optimization below.

All spectra in Eqs.~\eqref{eq_detector_thermal}--\eqref{eq_squid_imprecision_noise} are taken to be one-sided power spectral densities per unit frequency. Within this convention, we define the action-normalized SQUID noise product by
\begin{equation}
\epsilon
\equiv
\frac{\sqrt{S_{vv}S_{ii}}}{\omega_0}.
\label{eq_epsilon_squid}
\end{equation}
For the uncorrelated quantum-limited benchmark adopted here,
\begin{equation}
\epsilon=\hbar,
\qquad
S_{vi}=0,
\label{eq_sql_assumption_readout}
\end{equation}
so that
\begin{equation}
\sqrt{S_{vv}S_{ii}}
=
\hbar\omega_0.
\label{eq_sql_product_corrected}
\end{equation}
The factor $\omega_0$ is required for dimensional consistency when $S_{vv}$ and $S_{ii}$ are expressed as voltage- and current-noise power spectral densities per hertz.
The choice $\epsilon=\hbar$ is adopted as an ideal action-resolution benchmark in this effective, uncorrelated SQUID noise model. Its numerical normalization is specified by the assumed equivalent-input-noise convention and should not be interpreted as a device-independent quantum limit for an arbitrary SQUID implementation.

The SQUID noise impedance is defined as
\begin{equation}
Z_n
\equiv
\sqrt{\frac{S_{vv}}{S_{ii}}}.
\label{eq_noise_impedance_readout}
\end{equation}
At the target frequency, the value that balances the backaction and additive-noise contributions is
\begin{equation}
Z_{n,\rm match}
\equiv
\left|
\frac{
{\cal D}_3(\omega_0)
}{
{\cal A}_2(\omega_0)
}
\right|.
\label{eq_matched_noise_impedance}
\end{equation}
For optimization over a finite frequency band, we parameterize the actual noise impedance as
\begin{equation}
Z_n
=
\eta_n Z_{n,\rm match}
=
\eta_n
\left|
\frac{
{\cal D}_3(\omega_0)
}{
{\cal A}_2(\omega_0)
}
\right|,
\label{eq_eta_noise_impedance}
\end{equation}
where $\eta_n>0$ is treated as an adjustable readout parameter. The case $\eta_n=1$ gives exact backaction--additive-noise balance at $\omega_0$, whereas $\eta_n\neq1$ allows the worst added noise over the full analysis band to be reduced even when the two contributions are not equal at the center frequency.

Combining Eqs.~\eqref{eq_epsilon_squid} and~\eqref{eq_noise_impedance_readout}, the voltage- and current-noise spectra are
\begin{equation}
S_{vv}
=
\epsilon\omega_0Z_n,
\qquad
S_{ii}
=
\frac{\epsilon\omega_0}{Z_n}.
\label{eq_corrected_squid_psds}
\end{equation}
For the quantum-limited benchmark $\epsilon=\hbar$, this becomes
\begin{equation}
S_{vv}
=
\hbar\omega_0Z_n,
\qquad
S_{ii}
=
\frac{\hbar\omega_0}{Z_n}.
\label{eq_corrected_sql_squid_psds}
\end{equation}
These expressions replace formulas in which the factor $\omega_0$ is absent.

Equations~\eqref{eq_detector_thermal} and~\eqref{eq_readout_thermal} make the main experimental dependencies explicit. Since $\Lambda=NBA_{\rm eff}F(\theta,\psi)$, increasing the magnetic field, the effective pickup area, or the antenna projection increases the signal transfer function and therefore reduces the equivalent input-referred strain noise. For the detector thermal term, however, the turn-number dependence cancels once the detector inductance is written explicitly. For an ideal solenoidal pickup with effective area $A_{\rm eff}$ and length $\ell$,
\begin{equation}
L_d
=
\mu_0\frac{N^2A_{\rm eff}}{\ell},
\qquad
R_d
=
\frac{\omega_0L_d}{Q_d}.
\label{eq_Ld_solenoid_explicit}
\end{equation}
Substitution into Eq.~\eqref{eq_detector_thermal} gives
\begin{align}
S_d(\omega)
&=
\frac{4k_BT_d}{\omega^2(NBA_{\rm eff}F)^2}
\frac{\omega_0}{Q_d}
\mu_0\frac{N^2A_{\rm eff}}{\ell}
\nonumber\\
&=
\frac{
4\mu_0k_BT_d\omega_0
}{
Q_d\omega^2F^2B^2A_{\rm eff}\ell
}
\nonumber\\
&=
\frac{
4\mu_0k_BT_d\omega_0
}{
Q_d\omega^2F^2B^2V_{\rm sol}
},
\label{eq_detector_thermal_solenoid_volume}
\end{align}
where
\begin{equation}
V_{\rm sol}=A_{\rm eff}\ell
\end{equation}
is the effective solenoid volume. Thus the explicit factor $N^2$ in $L_d$ cancels the $N^2$ in the magnetic signal coefficient. In the ideal lumped-solenoid approximation, the strain-referred detector thermal floor is independent of turn number and is controlled instead by the magnetic field, quality factor, temperature, frequency, antenna projection, and effective solenoid volume. At the detector resonance,
\begin{equation}
S_d(\omega_0)
=
\frac{
4\mu_0k_BT_d
}{
Q_d\omega_0F^2B^2V_{\rm sol}
}.
\label{eq_detector_thermal_center_volume}
\end{equation}
This form shows directly that, for fixed field and quality factor, decreasing the system temperature and enlarging the active solenoid volume reduce the detector thermal strain floor, whereas adding turns at fixed volume does not by itself improve $S_d$.

The same cancellation should be understood as a property of the idealized loss model, not as a complete engineering rule. In a real coil, changing $N$ also changes winding volume, parasitic capacitance, self-resonance, residual loss, trapped flux, and achievable mutual coupling. The full three-loop optimization must therefore still be performed using the coupled-circuit transfer function. The useful design lesson is that the detector thermal floor is more directly tied to the high-field solenoid volume $V_{\rm sol}$ than to turn number alone, while $N$ remains an engineering parameter for realizing the required inductance, coupling, and resonance within practical constraints.

{The three-loop model} therefore contains two readout parameters that are absent when exact resonance and exact center-frequency noise matching are imposed: the readout detuning $\Delta f_A$ and the noise-impedance scale $\eta_n$. Together with $L_{A\mathrm{in}}$, $L_{A\mathrm{out}}$, $k_{\rm in}$, $L_s$, $L_i$, and $k_{\rm SQ}$, these parameters determine the frequency dependence of $S_{hh}^{\rm added}$. They can be optimized to minimize the maximum added-noise amplitude spectral density over a finite band around $f_0$, while the detector thermal term remains fixed by $B$, $V_{\rm sol}$, $Q_d$, $T_d$, and the antenna factor.

The main advantage of the electromagnetic implementation is that the resonant frequency, coupling, readout detuning, and readout noise impedance can be engineered electrically while retaining a compact pickup structure inside a high-field magnet. This makes the detector smaller and more tunable than a ton-scale acoustic antenna, while still exploiting resonant enhancement and low-noise SQUID or quantum-limited readout.

\section{Prototype superconducting LC resonator}
\label{sec_prototype_lc}

To achieve maximum sensitivity in the proposed {gravitational-wave detector}, the intrinsic quality factors of the LC resonators in the detector and the readout circuit must be as high as possible, with a target of $Q\sim10^7$ at $14\text{ T}$.

Several dissipation mechanisms can limit the quality factor of an LC resonator, including dielectric losses in the capacitor and in the parasitic capacitance of the coil; magnetic losses within the coil formers; dissipation due to eddy currents induced in metallic components near the coil; ohmic losses in the resonator circuit unless the circuit is fully superconducting; dissipation caused by the motion induced in fluxons trapped within the superconductors; and dissipation introduced by the readout amplifier.

Through a proper choice of materials and fabrication techniques, and by employing a SQUID as the readout amplifier, intrinsic quality factors on the order of $10^6$ can be achieved for a resonator operating at approximately 1 kHz.

\begin{figure*}[t]
\centering
\includegraphics[width=\linewidth]{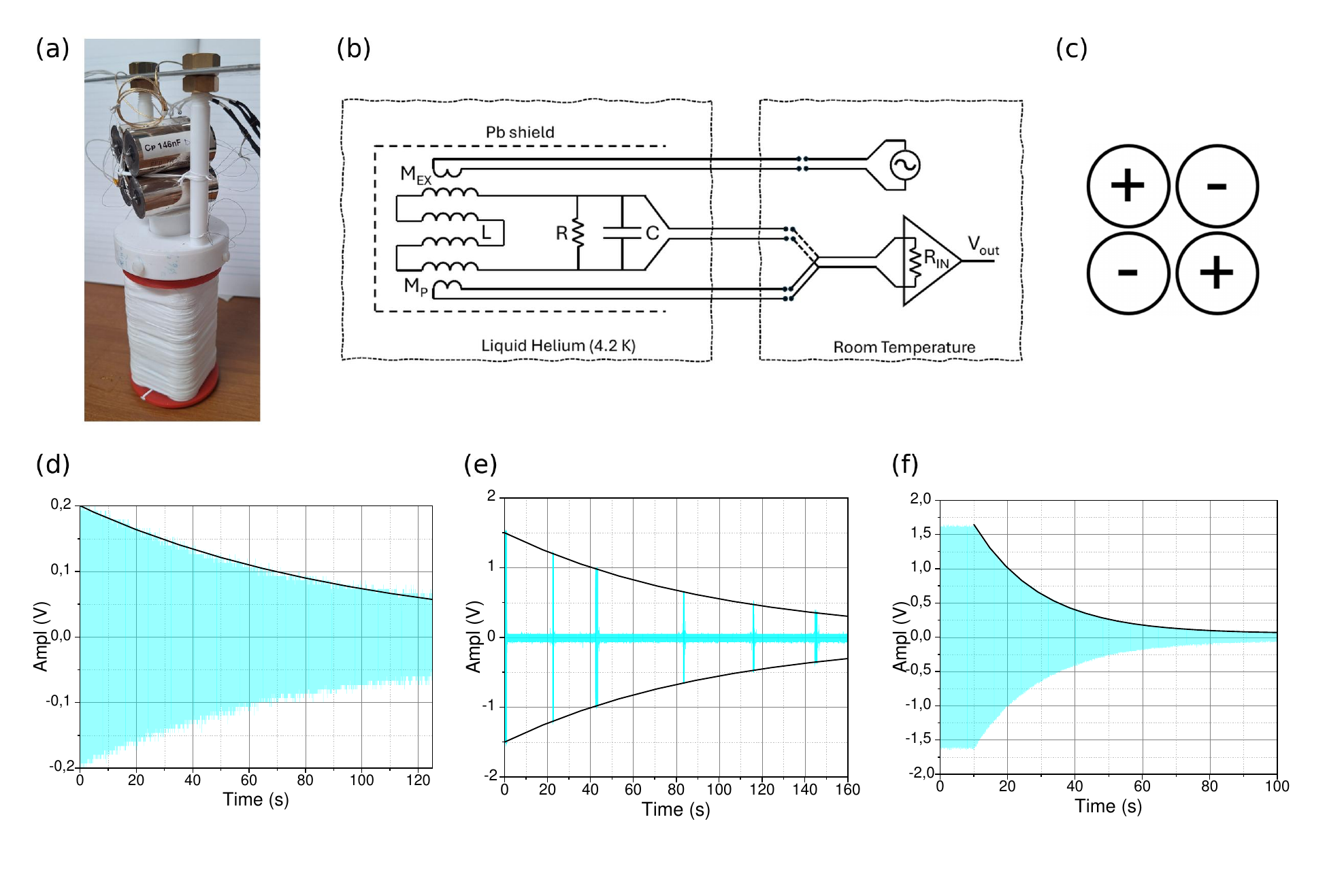}
\caption{
(a) Photograph of the LC resonator consisting of four series-connected solenoids forming inductor L and four commercial capacitors in parallel forming capacitor C. The LC resonator is operated immersed in liquid helium, both with and without a superconducting lead shield.
(b) Schematic representation of the experimental setup used for the measurements. R represents the resonator losses, $M_{ex}$ the coupling between the excitation coil and inductor L, $M_p$ the weak coupling between the pickup coil and inductor L, and $R_{in}$ the input impedance of the SR560 amplifier used to send the signal to the oscilloscope.
(c) Schematic representation of the positions and winding orientations of the four solenoids.
(d) Signal decay and best fit for the resonator with a superconducting lead shield, acquired via the coupling $M_p$ between the pickup and inductor L.
(e) Signal decay sampled over short intervals and best fit for the resonator with a superconducting lead shield, acquired via direct coupling between the amplifier and resonator.
(f) Signal decay and best fit for the resonator without a superconducting lead shield, acquired via the coupling $M_p$ between the pickup and inductor L.
}
\label{fig_LC_Characterization}
\end{figure*}

{To provide an experimental benchmark for the proposed detector concept, a low-loss} LC resonator was built as shown in Figure~\ref{fig_LC_Characterization}(a) with the following specifications. The inductor $L$ in Figure~\ref{fig_LC_Characterization}(b) consists of four identical series-connected solenoids, realized by winding a single layer of Nb-47wt\%Ti wire, 75 $\mu$m in bare diameter and 100 $\mu$m including the Formvar insulating varnish, onto PVC formers. A single-layer winding scheme was selected to minimize parasitic capacitance, thereby minimizing the associated dielectric losses.

The solenoids have a length of 88 mm and a diameter of 24 mm, arranged in an alternating orientation as shown in Figure~\ref{fig_LC_Characterization}(c). While this specific coil arrangement prevents the detection of the actual gravitational wave signal, it serves in these preliminary tests to localize the magnetic field as much as possible, reducing dissipation induced in nearby metallic components, such as the capacitors and the metallic walls of the liquid-helium dewar used for the test.

The capacitor bank had a measured room-temperature capacitance of $C = 592\text{ nF}$ and was constructed by connecting four Eurofarad TA72 Teflon capacitors in parallel, each with a nominal capacitance of 0.15 $\mu$F and a maximum rated voltage of 400 V. The coil and capacitors are supported by a Teflon structure. Magnetic losses in this support and in the solenoids' PVC formers are negligible.

The measurement setup is shown in Figure~\ref{fig_LC_Characterization}(b). An amplifier (Stanford Research Systems SR560) with an input resistance of $100\text{ M}\Omega$ can be connected either directly to the LC resonator or to a weakly coupled pickup coil ($M_p$) wound around one of the four solenoids. The resonator is excited via inductive coupling ($M_{\text{ex}}$) using a second superconducting coil and then allowed to ring down freely to measure the decay time constant and determine the quality factor. The output signal from the amplifier was acquired by an oscilloscope. The quality factors were measured with and without a cylindrical superconducting lead shield to quantify dissipation associated with the metallic inner walls of the liquid helium dewar.

In the first series of measurements, performed with the superconducting shield, the resonance frequency was $f_0 = 1534\text{ Hz}$. Assuming that the capacitance of the four parallel capacitors retained its room-temperature value at $4.2\text{ K}$ in liquid helium ($C = 592\text{ nF}$), the inductance of the coil can be estimated as $L \approx 18\text{ mH}$.

From the decay measurements acquired via the pickup coil as shown in Figure~\ref{fig_LC_Characterization}(d), an intrinsic quality factor of $Q_i = 5.0 \times 10^5$ was determined. Given the extremely weak coupling between the pickup and the resonator coil, the influence of the amplifier input impedance on this $Q$ value is expected to be negligible; the measured value can therefore be taken as the intrinsic $Q$ of the resonator.

This assumption was subsequently validated by specific measurements performed with direct coupling between the amplifier and the LC resonator. In these tests, the signal line was connected to the amplifier only for short intervals during the ring-downs as shown in Figure~\ref{fig_LC_Characterization}(e), allowing the signal to decay freely for most of the time without loading from the amplifier's input resistance. A fit of the data acquired with this method confirmed a quality factor of $Q = 5.0 \times 10^5$.

Conversely, when the amplifier remained continuously connected to the resonator during a decay, the measured quality factor dropped to $Q_m \approx 2.5 \times 10^5$. This result is in good agreement with theoretical expectations. Denoting by $R$ the equivalent parallel resistance accounting for the intrinsic losses of the resonator, we have:
\begin{equation}
Q_i = 2\pi f_0 C R = 5.0 \times 10^5
\label{eq_prototype_intrinsic_Q}
\end{equation}
yielding $R \approx 88\text{ M}\Omega$. The quality factor associated solely with the amplifier's input resistance ($R_a = 100\text{ M}\Omega$) is given by:
\begin{equation}
Q_a = 2\pi f_0 C R_a \approx 5.7 \times 10^5
\end{equation}
The total expected quality factor resulting from both dissipation sources is:
\begin{equation}
Q = \left( \frac{1}{Q_i} + \frac{1}{Q_a} \right)^{-1} \approx 2.66 \times 10^5
\label{eq_prototype_loaded_Q}
\end{equation}
which is in good agreement with the measured value $Q_m$. A second series of measurements was conducted with the resonator operating without the superconducting shield. In this setup, eddy current losses induced in the metallic walls of the liquid helium dewar added to the intrinsic losses, reducing the overall quality factor to approximately $1.0 \times 10^5$ as shown in Figure~\ref{fig_LC_Characterization}(f).

{The resonator quality factor could potentially be improved to $Q\sim10^7$.} Previous superconducting LC resonators {for the AURIGA experiment} have demonstrated quality factors exceeding $10^6$ at frequencies comparable to our experiment~\cite{Falferi1994HighQLC,Bonaldi1998HighQLC}, while recent developments for low-mass axion searches have achieved $Q\simeq2.1\times10^6$~\cite{Kolevatov2026HighQLumped}. Furthermore, a NbTi superconducting radio-frequency cavity maintained $Q_0=4.5\times10^5$ at 2~T~\cite{Alesini2019SuperconductingCavity}. Since vortex-motion-induced resistance decreases at frequencies below the depinning frequency, substantial intrinsic degradation of the NbTi resonator performance at 2~T is not necessarily expected in the kHz regime. Any observed reduction in $Q$ may therefore also arise from resonator design, shield design, or other experimental factors rather than intrinsic conductor losses alone. In the longer term, high-temperature superconducting (HTS) wires may enable $Q\sim10^7$ operation at 14~T~\cite{Utschick2021thesis}, although this remains to be experimentally demonstrated.

\section{Projected sensitivity}
\label{sec_projected_sensitivity}

\begin{figure*}[t]
\centering
\includegraphics[width=0.9\textwidth]{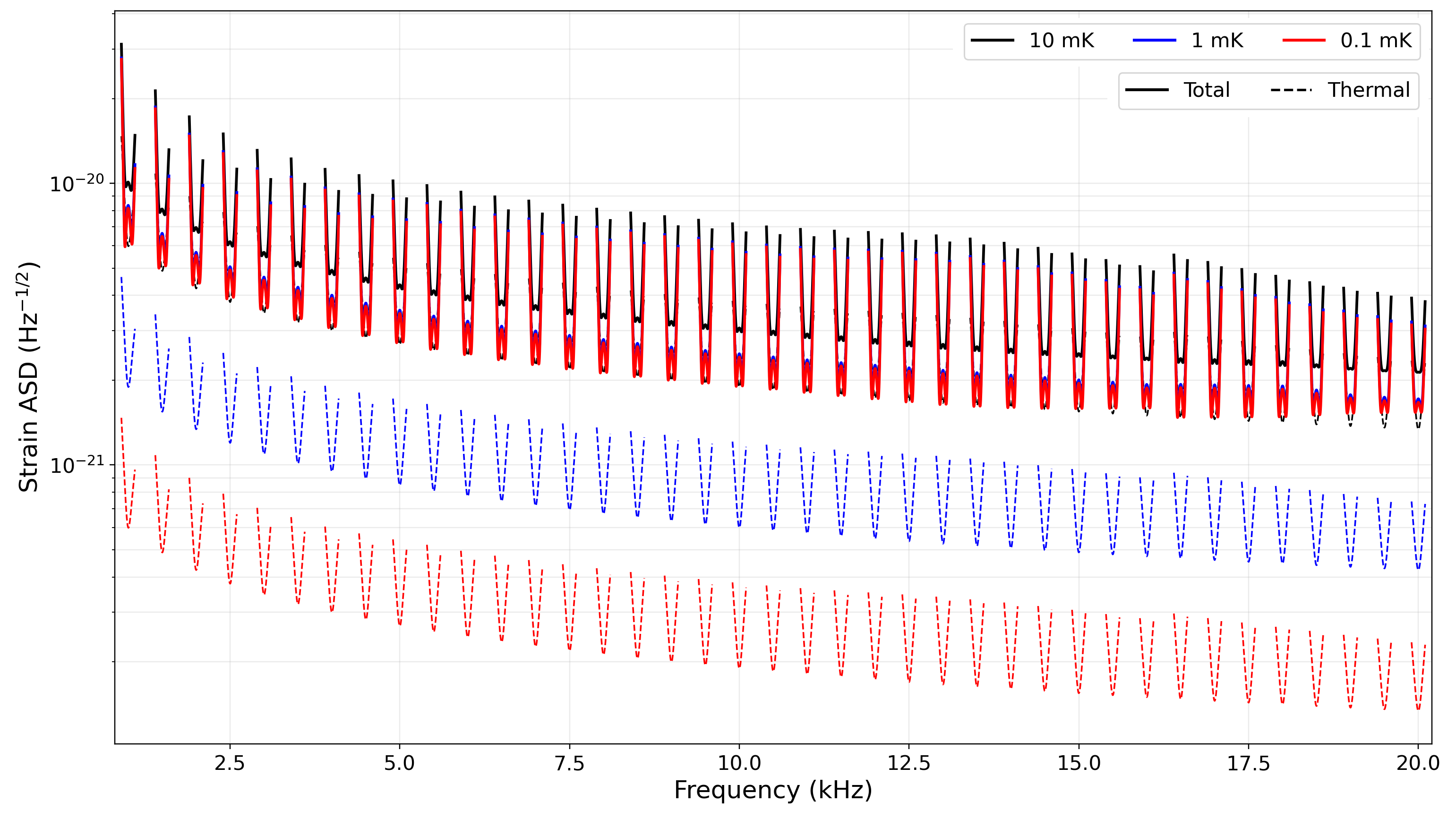}
\caption{
Projected input-referred strain-noise amplitude spectral density for {the} $Q_d=Q_A=10^7$ reference circuits. Each circuit is optimized once at $T_{\rm ref}=10~{\rm mK}$ and then evaluated without parameter readjustment at lower temperatures. Solid curves show the total ASD and thin dashed curves show the thermal contribution $(S_d+S_A)^{1/2}$. The displayed temperatures are $10~{\rm mK}$ (black), $1~{\rm mK}$ (blue), and $0.1~{\rm mK}$ (red). For each center frequency only the local optimized response is plotted, so the figure represents a collection of narrow-band tunable configurations rather than one simultaneous broadband detector response.
}
\label{fig_temperature_overlay}
\end{figure*}

We now evaluate the projected strain sensitivity of the three-loop detector--readout--SQUID chain described in Sec.~\ref{sec_readout}. The purpose of this section is not to claim a measured gravitational-wave sensitivity, but to define a transparent benchmark for the optimized coupled circuit and to identify which experimental assumptions control the reach. The calculation uses the input-referred strain-noise model of Eq.~\eqref{eq_total_shh}, including detector thermal noise, readout thermal noise, SQUID backaction, and SQUID additive noise. The plotted quantity is {the one-sided strain-noise amplitude spectral density (ASD),}
\begin{equation}
{S_{hh}^{1/2}(\omega)}
=
\left[
S_d(\omega)
+
S_A(\omega)
+
S_{\rm ba}(\omega)
+
S_{\rm add}(\omega)
\right]^{1/2}.
\label{eq_projected_total_asd}
\end{equation}
The $Q=10^7$ values used below represent targeted future detector and readout performance, rather than the measured quality factor of the present prototype.
Unless stated otherwise, the detector and readout quality factors are fixed to
\begin{equation}
Q_d=Q_A=10^7,
\end{equation}
the detector pickup coil has $N=1000$ turns, and the temperature set is
\begin{equation}
T_d=T_A
=
10,\ 1,\ 0.1~{\rm mK}.
\label{eq_temperature_set_revised}
\end{equation}
The center-frequency grid is
\begin{equation}
\omega_0
=
2\pi\times(1.0,\ 1.5,\ 2.0,\ldots,\ 20.0)~{\rm kHz},
\label{eq_projected_target_frequencies}
\end{equation}
giving 39 frequency-specific reference circuit settings for $Q=10^7$, obtained by warm-start frequency continuation with a fixed input-coil geometry. The lowest-temperature case, $0.1~{\rm mK}=100~\mu{\rm K}$, is an effective circuit-mode-temperature projection and not a demonstrated operating point of the complete high-field apparatus.

All analytical spectra and optimization conditions in this section are expressed in angular frequency $\omega$; frequency values and figure-axis offsets quoted in Hz are the corresponding ordinary frequencies, related by $\omega=2\pi f$.

\subsection{Optimization strategy and constraints}
\label{subsec_optimization_strategy}

For each target frequency, the detector capacitance is chosen so that the detector resonance is centered at
\begin{equation}
C_d
=
\frac{1}{\omega_0^2L_d}.
\label{eq_projected_Cd}
\end{equation}
The readout resonance is not forced to coincide exactly with the detector resonance. Instead,
\begin{equation}
\omega_A
=
\omega_0+\Delta\omega_A,
\qquad
C_A
=
\frac{1}{
\omega_A^2
\left(
L_{A{\rm in}}+L_{A{\rm out}}
\right)
},
\label{eq_projected_CA}
\end{equation}
where the detuning $\Delta\omega_A$ is optimized together with the other circuit parameters. The input inductance $L_{A\mathrm{in}}$ and coupling $k_{\rm in}$ are fixed by a discrete, centered coaxial input-coil geometry {($L_{A\mathrm{in}}\simeq0.35~\mu\mathrm{H}$ and $k_{\rm in}\simeq0.34$)}. The remaining six parameters are optimized using frequency continuation; the following vector retains the original circuit parameter notation:
\begin{equation}
\bm p
=
\left(
L_{A{\rm in}},
L_{A{\rm out}},
k_{\rm in},
L_s,
L_i,
k_{\rm SQ},
\eta_n,
\Delta\omega_A
\right).
\label{eq_projected_parameter_vector}
\end{equation}
For each target $\omega_0$ at $Q_d=Q_A=10^7$, the six adjustable entries are optimized once at the reference temperature
\begin{equation}
T_{\rm ref}=10~{\rm mK}.
\end{equation}
The same optimized circuit settings are then copied without further adjustment to the $1$, $0.1$, and $0.01~{\rm mK}$ evaluations. Thus the temperature comparison describes cooling of a fixed physical circuit rather than a separate re-optimization at every temperature.

The physical minimax target is the worst added-noise ASD in the central $100~{\rm Hz}$ band,
\begin{equation}
\min_{\bm p}
\left[
\max_{|\omega-\omega_0|\le 2\pi\times50~{\rm Hz}}
\left(
S_{hh}^{\rm added}(\omega;\bm p,T_{\rm ref})
\right)^{1/2}
\right],
\label{eq_revised_minimax_objective}
\end{equation}
where
\begin{equation}
S_{hh}^{\rm added}
=
S_A+S_{\rm ba}+S_{\rm add}.
\end{equation}
The numerical implementation uses a smooth approximation to the worst-case term together with penalties enforcing the band-shape requirements. The $800~\mu\mathrm{H}$ upper bound on $L_{A\mathrm{out}}$ is imposed, and successive target frequencies are initialized from neighboring optimized settings. This seeded local continuation does not certify global optimality. The mechanical feasibility of the SQUID output coupling, parasitic capacitance, and loaded quality factors is not established by this lumped-element calculation. In contrast to a prescribed mode splitting, the locations of the two sensitivity minima are not fixed in advance. The design is required to contain one interior local minimum on each side of $f_0$, with nearly equal depths,
\begin{equation}
\left|
\ln
\frac{
S_{h,L}^{\rm added\,1/2}
}{
S_{h,R}^{\rm added\,1/2}
}
\right|
\le 10^{-3},
\label{eq_equal_minima_constraint}
\end{equation}
and the added-ASD ripple must satisfy
\begin{equation}
\frac{
\displaystyle
\max_{|\omega-\omega_0|\le2\pi\times50~{\rm Hz}}
S_{hh}^{\rm added\,1/2}(\omega)
}{
\displaystyle
\min_{|\omega-\omega_0|\le2\pi\times50~{\rm Hz}}
S_{hh}^{\rm added\,1/2}(\omega)
}
\le\sqrt{2}.
\label{eq_revised_ripple_constraint}
\end{equation}
The latter corresponds to a $3~{\rm dB}$ limit in the added-noise PSD.
The equal-minima and 3-dB flatness criteria are heuristic
design constraints introduced to obtain an approximately
uniform added-noise amplitude spectral density over the
chosen 100-Hz analysis band, rather than fundamental
physical requirements of the detector.

\begin{table}[t]
\caption{
Optimization domain for the geometry-constrained, $800~\mu\mathrm{H}$, $Q=10^7$ projected-sensitivity calculation.
}
\label{tab_optimization_constraints}
\scriptsize
\begin{ruledtabular}
\begin{tabular}{lll}
Parameter & Range & Role \\
\hline
$L_{A{\rm in}}$ &
$0.35~\mu{\rm H}$ (fixed) &
fixed input
\\
$L_{A{\rm out}}$ &
$10^{-8}~{\rm H}$--$800~\mu{\rm H}$ &
readout output
\\
$k_{\rm in}$ &
$0.34$ (fixed) &
fixed input coupling
\\
$L_s$ &
$0.10$--$1.0~\mu{\rm H}$ &
SQUID-loop 
\\
$L_i$ &
$0.50$--$3.0~\mu{\rm H}$ &
SQUID input
\\
$k_{\rm SQ}$ &
$0.10$--$0.60$ &
readout--SQUID coupling
\\
$\eta_n$ &
$10^{-3}$--$10^3$ &
noise-impedance scale
\\
$\Delta\omega_A$ &
$-2\pi\times250$--$2\pi\times250~{\rm rad\,s^{-1}}$ &
readout detuning
\\
Analysis band &
$|\omega-\omega_0|\le2\pi\times50~{\rm Hz}$ &
flatness band
\\
Added-ASD ripple &
$\le\sqrt{2}$ &
$3~{\rm dB}$ PSD limit
\end{tabular}
\end{ruledtabular}
\end{table}
All 39 numerically verified reference solutions for $Q=10^7$ satisfy the optimization bounds, the $3~{\rm dB}$ added-noise flatness condition, the equal-minima tolerance, and the requirement of two interior local minima at $10~{\rm mK}$. Because the lower-temperature spectra are evaluated with the parameters fixed, the equal-minima condition is not reimposed after cooling. The $3~{\rm dB}$ flatness condition remains satisfied for all 39 circuits at each lower temperature, while the detailed local-minimum structure changes as the thermal terms decrease relative to the temperature-independent SQUID terms. This distinction is important: the lower-temperature curves are fixed-circuit projections, not independently optimized solutions.
\begin{table}[t]
\caption{
Fixed detector parameters used in {the} projected-sensitivity calculation.
}
\label{tab_fixed_projected_parameters}
\scriptsize
\begin{ruledtabular}
\begin{tabular}{lll}
Parameter & Symbol & Value \\
\hline
Magnetic field & $B$ & $14~{\rm T}$ \\
Turns & $N$ & $1000$ \\
Diameter & $d$ & $0.10~{\rm m}$ \\
Length & $\ell$ & $0.20~{\rm m}$ \\
Effective area & $A_{\rm eff}$ & $7.854\times10^{-3}~{\rm m^2}$ \\
Inductance & $L_d$ & $4.93\times10^{-2}~{\rm H}$ \\
Signal coefficient & $NBA_{\rm eff}$ & $1.10\times10^{2}~{\rm Wb}$ \\
Detector $Q$ & $Q_d$ & $10^7$ \\
Readout $Q$ & $Q_A$ & $10^7$ \\
Reference temperature & $T_{\rm ref}$ & $10~{\rm mK}$ \\
Evaluation temperatures & $T_d=T_A$ & ${10,1,0.1~{\rm mK}}$ \\
\end{tabular}
\end{ruledtabular}
\end{table}
The input geometry is kept fixed throughout the frequency scan; the SQUID-side inductance $L_{A\mathrm{out}}$ is bounded by $800~\mu\mathrm{H}$. All numerical sensitivity projections assume the same benchmark GW-to-flux coefficient as in {this work} and do not establish a mechanically calculated coupling. The fixed detector and cryogenic parameters used in the benchmark are listed in Table~\ref{tab_fixed_projected_parameters}.
As a representative example, the $\omega_0=2\pi\times5~{\rm kHz}$ reference optimization gives
\begin{equation}
\begin{aligned}
\eta_n&=11.34, \quad \Delta\omega_A=-2\pi\times3.92~{\rm rad\,s^{-1}},\\
Z_n&=0.53~\Omega.
\end{aligned}
\label{eq_5k_reference_matching}
\end{equation}
The two added-noise minima occur at
\begin{equation}
\omega-\omega_0=-2\pi\times33.30,
\qquad
+2\pi\times34.15~{\rm rad\,s^{-1}},
\end{equation}
and the added-ASD ripple  over $|\omega-\omega_0|\le2\pi\times50~{\rm Hz}$ is $1.08789$. This illustrates why neither exact readout tuning nor a prescribed $2\pi\times(\pm25~{\rm Hz})$ angular-frequency splitting is imposed in {the} optimization. Although $N=1000$ is retained as the circuit-design value, the ideal lumped-solenoid detector thermal floor does not improve merely by increasing $N$ at fixed volume. As shown in Eq.~\eqref{eq_detector_thermal_center_volume}, the $N^2$ scaling of the signal coefficient is cancelled by the $N^2$ scaling of the solenoid inductance.

\subsection{Noise decomposition and quantum-readout limit}
\label{subsec_noise_decomposition}

%\begin{figure}[t]
%\centering
%\includegraphics[width=\linewidth]{figure/cc.png}
%\caption{
%{
%Center-frequency projected strain sensitivity for $Q_d=Q_A=10^7$. Solid curves show the total center ASD, while dashed curves show the combined detector and readout thermal contribution $\left(S_d+S_A\right)^{1/2}$ at $T_d=T_A=10$, $1$, and $0.1~{\rm mK}$. The black, blue, and red curves correspond to $10~{\rm mK}$, $1~{\rm mK}$, and $0.1~{\rm mK}$, respectively.
%}
%}
%\label{fig_temperature_center}
%\end{figure}

\begin{figure}[t]
\centering
\includegraphics[width=\linewidth]{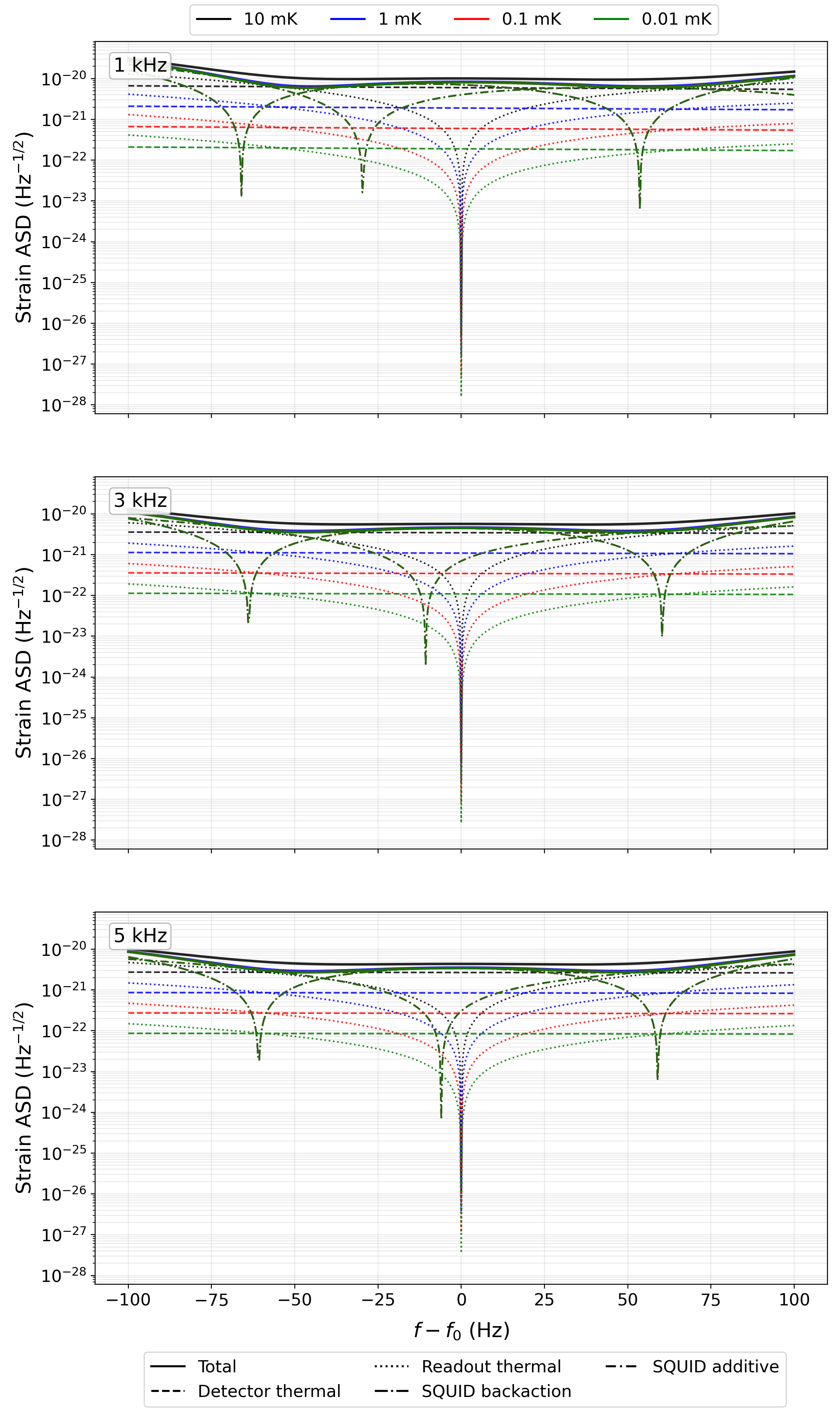}
\caption{
Full local component decomposition around the optimized $1$, $3$, and $5~{\rm kHz}$ targets for $Q_d=Q_A=10^7$. Colors denote the fixed-circuit evaluation temperatures $10$, $1$, $0.1$, and $0.01~{\rm mK}$, while line style denotes the total, detector thermal, readout thermal, SQUID backaction, and SQUID additive-noise contributions. The horizontal axis is the offset $f-f_0$ over a $\pm100~{\rm Hz}$ window. The reference circuits are optimized at $10~{\rm mK}$; the lower-temperature curves use exactly the same parameters.
}
\label{fig_component_decomposition_135kHz}
\end{figure}

Figure~\ref{fig_temperature_overlay} shows the wide frequency scan for the fixed $10~{\rm mK}$ reference circuits. The thermal contribution decreases as $T^{1/2}$, but the total ASD does not follow the same scaling because the SQUID backaction and additive noise remain fixed when the circuit parameters and the action resolution $\epsilon=\hbar$ are held constant. Consequently, cooling produces only a modest improvement once the SQUID-added floor dominates. The improvement between $10~{\rm mK}$ and $100~\mu{\rm K}$ is only a factor of approximately $1.24$--$1.28$ in center ASD over the $1$--$20~{\rm kHz}$ grid, rather than the thermal-only factor $\sqrt{100}$. The reason is that the total sensitivity is already close to the fixed SQUID-added floor.
Table~\ref{tab_5kHz_projected_components} gives the numerical decomposition at $5~{\rm kHz}$. The readout thermal contribution is negligible in this optimized example. The combined SQUID-added center ASD is
\begin{equation}
\left(
S_{\rm ba}+S_{\rm add}
\right)^{1/2}
=
3.44\times10^{-21}~{\rm Hz}^{-1/2},
\end{equation}
independent of temperature. The total center ASD therefore decreases only from
$4.36\times10^{-21}~{\rm Hz}^{-1/2}$ at $10~{\rm mK}$ to
$3.45\times10^{-21}~{\rm Hz}^{-1/2}$ at $100~\mu{\rm K}$, a factor of approximately $1.26$.

\begin{table*}[t]
\caption{
{
Representative center-frequency noise components at
$\omega_0=2\pi\times5~{\rm kHz}$ for {the} $Q_d=Q_A=10^7$
fixed-circuit temperature sweep. All entries are amplitude spectral
densities in units of $10^{-21}~{\rm Hz}^{-1/2}$.
}
}
\label{tab_5kHz_projected_components}
\centering
\small
\begingroup
\setlength{\tabcolsep}{7pt}
\renewcommand{\arraystretch}{1.12}
\begin{ruledtabular}
\begin{tabular}{cccccc}
$T_d=T_A$ &
$S_d^{1/2}$ &
$S_A^{1/2}$ &
$S_{\rm ba}^{1/2}$ &
$S_{\rm add}^{1/2}$ &
{$S_{hh}^{1/2}$}
\\
\hline
$10~{\rm mK}$ & $2.68$ & $1.12e-05$ & $3.43$ & $0.30$ & $4.36$ \\
$1~{\rm mK}$ & $0.85$ & $3.53e-06$ & $3.43$ & $0.30$ & $3.54$ \\
$0.1~{\rm mK}$ & $0.27$ & $1.12e-06$ & $3.43$ & $0.30$ & $3.45$ \\
\end{tabular}
\end{ruledtabular}
\endgroup
\end{table*}
Figure~\ref{fig_component_decomposition_135kHz} shows how the local line shape evolves as the thermal terms are reduced. The detector thermal ASD is nearly constant over the narrow local band and scales as $T^{1/2}$. The readout thermal term also scales as $T^{1/2}$ but is much smaller in the representative optimized circuits. By contrast, the SQUID backaction and additive noise retain the same amplitudes at all temperatures and have complementary frequency dependence. The total low-temperature spectrum therefore approaches the SQUID-added structure rather than continuing to decrease as $T^{1/2}$.

In the classical thermal-noise limit, $k_{\rm B}T\gg\hbar\omega$, the thermal terms scale approximately as
\begin{equation}
S_{h,{\rm th}}^{1/2}\propto T^{1/2}.
\label{eq_temperature_scaling}
\end{equation}
{
For the $1$--$20~{\rm kHz}$ frequency range,
\begin{equation}
\frac{\hbar\omega}{k_{\rm B}}
\simeq
0.05\text{--}1~\mu{\rm K}.
\end{equation}
Thus the lowest effective temperature used here, $100~\mu{\rm K}$, remains above the single-quantum temperature scale. Nevertheless, it is far below a conventional dilution-refrigerator mixing-chamber benchmark and should be interpreted as an effective temperature of the relevant circuit modes, not as the temperature of the entire magnet, shield, wiring, and SQUID assembly. Feedback cooling of AURIGA normal modes from a $4.2~{\rm K}$ bath to $0.17~{\rm mK}$ demonstrates that very low effective mode temperatures can be produced in a resonant detector~\cite{Vinante2008zz}. Nuclear-demagnetization and microkelvin nanoelectronic platforms also demonstrate access to sub-mK electronic temperatures~\cite{Nyeki2022microK,Sarsby2020500microkelvin}, but these results do not establish operation of the complete proposed high-field detector at the same effective temperature.
}

{
Following the standard quantum-noise treatment of linear amplifiers and low-backaction SQUID readout, the SQUID is modeled as an uncorrelated phase-preserving quantum-limited amplifier in the one-sided spectral-density convention adopted here~\cite{Clerk2010rmp,ClarkeBraginski2004,Falferi1998apl},
\begin{equation}
S_{vi}=0,
\qquad
\frac{\sqrt{S_{vv}S_{ii}}}{\omega_0}
=
\epsilon
=
\hbar.
\label{eq_sql_assumption_projected}
\end{equation}
Its noise impedance is
\begin{equation}
Z_n
=
\eta_n
\left|
\frac{
{\cal D}_3(\omega_0)
}{
{\cal A}_2(\omega_0)
}
\right|,
\label{eq_Zn_projected}
\end{equation}
where the ratio of voltage-noise backaction to current-noise additive contribution is expressed through an effective amplifier noise impedance, as in the standard linear-detector formalism~\cite{Clerk2010rmp}. The parameter $\eta_n$ is included in the finite-band optimization. Therefore, {the} calculation does not assume exact center-frequency matching, $\eta_n=1$, but permits the backaction--additive-noise balance to shift when doing so reduces the worst added noise over the entire $100~{\rm Hz}$ band. At the representative $5~{\rm kHz}$ point, $\eta_n=11.34$, and the center-frequency backaction consequently exceeds the center-frequency additive contribution, as shown in Table~\ref{tab_5kHz_projected_components}.

For the base geometry with $Q_d=Q_A=10^7$, reducing the effective mode temperature below $10~{\rm mK}$ does not produce orders-of-magnitude sensitivity improvement because the optimized SQUID-added contribution is already comparable to or larger than the detector thermal noise. As shown in Fig.~\ref{fig_temperature_overlay}, the projected center sensitivity remains above $10^{-21}~{\rm Hz}^{-1/2}$ throughout the $1$--$20~{\rm kHz}$ grid, even at $100~\mu{\rm K}$. Reaching substantially below this level in the same frequency range would require improved signal transduction, such as a larger active high-field volume, together with a readout strategy that reduces the relevant added-noise quadrature. Possible approaches include exploiting imprecision--backaction correlations and phase-sensitive amplification~\cite{Caves1982,Clerk2010rmp}, or implementing a genuinely backaction-evading measurement using two-quadrature or quantum-upconversion techniques~\cite{Hertzberg2010,Suh2014,Kuenstner2025RQU}. The curves presented here should therefore be interpreted as fixed-circuit, {uncorrelated standard-quantum-limit (SQL)-like projections} under the stated $Q=10^7$, geometry, and effective-temperature assumptions, and not as a demonstrated phase-sensitive or backaction-evading sensitivity.

\section{Magnet requirements}
\label{sec_magnet_requirement}

{The LC-circuit/SQUID-readout model and the projected-sensitivity calculations} in Sec.~\ref{sec_readout} and Sec.~\ref{sec_projected_sensitivity} assume that the static field $B$ is sufficiently stable during the measurement and that magnetic backgrounds generated by the magnet system are below the input-referred strain noise. The magnet is therefore both a source of signal gain and a potential source of background.
Magnetic-field fluctuations coupled to the pickup loop produce an equivalent strain noise. For the idealized pickup geometry, keeping this contribution below the target sensitivity requires
\begin{equation}
\left(\frac{\delta B}{B}\right)_{\rm req}
\lesssim S_{hh,\rm min}^{1/2}
\label{eq_db_over_b_hmin_requirement}
\end{equation}
where both amplitudes are evaluated over the same measurement bandwidth. In spectral terms, a target strain-noise amplitude spectral density of $S_{hh}^{1/2}\sim10^{-23}~{\rm Hz}^{-1/2}$ requires approximately $S_{\delta B/B}^{1/2}\lesssim10^{-23}~{\rm Hz}^{-1/2}$ in the analysis band. This condition applies to residual magnetic fluctuations after stabilization, spectral filtering, and common-mode rejection, rather than to the raw DC drift of the magnet. Slow persistent-mode decay and magnetic noise in the kHz detection band must therefore be treated as distinct experimental backgrounds.

\subsection{Suspension and vibration-induced field noise}
\label{subsec_magnet_suspension}

Relative motion between the pickup circuit and the magnet in a
static magnetic-field gradient can produce a spurious flux signal.
For small displacements, the leading contribution from relative
translation may be estimated as
\begin{equation}
\left(\frac{\delta B}{B}\right)_{\rm vib}
\simeq
g_B x_{\rm rel},
\qquad
g_B\equiv\frac{|\nabla B|}{B},
\label{eq_gradient_vibration_coupling}
\end{equation}
where $x_{\rm rel}$ is the relative displacement projected along
the relevant field gradient. This expression provides an
order-of-magnitude estimate for translational coupling; changes
in pickup area, orientation, or magnetic-field distribution can
produce additional flux perturbations. The magnet specification
must therefore include a measured field-gradient map and the
mechanical transfer functions of the magnet, pickup, shield,
and associated support structures.

The AURIGA cryogenic suspension provides a useful benchmark
for mechanical isolation. Its upgraded suspension was designed
to achieve at least $-240~{\rm dB}$ of vibration attenuation near
$1~{\rm kHz}$, corresponding to an amplitude transfer factor $H_{\rm rel}$
of order $10^{-12}$ \cite{Bignotto2005rsi}.
This value represents a design benchmark rather than a
demonstrated isolation factor for the present detector.
For environmental displacement noise $S_{x,\rm env}$, the
fractional magnetic-field noise associated with relative
translation is approximately
\begin{equation}
S_{\delta B/B}^{1/2}(f)
\simeq
g_B |H_{\rm rel}(f)|
S_{x,\rm env}^{1/2}(f),
\label{eq_vibration_asd}
\end{equation}
where $H_{\rm rel}$ is the transfer function from environmental
motion to relative magnet--pickup displacement.

As an illustrative benchmark, taking
$g_B=10^{-4}~{\rm m}^{-1}$,
$|H_{\rm rel}|=10^{-12}$, and
$S_{x,\rm env}^{1/2}=10^{-9}~{\rm m}/\sqrt{\rm Hz}$
gives
$S_{\delta B/B}^{1/2}\simeq10^{-25}/\sqrt{\rm Hz}$,
below a representative optimal-orientation strain-noise
target of $10^{-23}/\sqrt{\rm Hz}$.
These assumed parameters are not established for the proposed
detector, and the estimate does not include other
vibration-induced flux couplings. The integrated noise over
a bandwidth $\Delta f$ is obtained from the corresponding
power spectral density and scales as $\sqrt{\Delta f}$
only when the spectrum is approximately flat.

In a linear time-invariant system, low-frequency excitation
does not directly generate spectral components in the kHz
detection band. However, nonlinear mechanical contacts,
cable microphonics, motion-dependent magnetic coupling,
and time-dependent circuit parameters can produce frequency
mixing or modulation. The magnet qualification should
therefore include a field-gradient map, mechanical transfer
function measurements covering the search band, and
controlled low-frequency excitation tests while monitoring
the $1$--$20~{\rm kHz}$ magnetic-flux spectrum.

Elongated suspended components, including support rods,
shields, cables, and coil formers, may also exhibit internal
mechanical resonances. For an isolated, weakly damped mode
with resonance frequency $f_n$ and quality factor $Q_n$,
the normalized force-to-displacement response is
\begin{equation}
|H_n(f)|
\simeq
\left[
\left(1-\frac{f^2}{f_n^2}\right)^2
+
\left(\frac{f}{f_n Q_n}\right)^2
\right]^{-1/2}.
\end{equation}
The resonant response reaches a factor of order $Q_n$,
with a linewidth $\Delta f_n\simeq f_n/Q_n$.
Measurements on isolated specimens have reported
$Q\simeq2\times10^3$ for SUS304 stainless steel rods,
$Q\simeq5\times10^4$ for duralumin rods, and
$Q>10^7$ for a cryogenic sapphire-fiber bending mode
\cite{Tsuchida2024QFactor,Uchiyama2000SapphireFiberQ}.
These values are material- and geometry-dependent examples
and do not establish the quality factors of the assembled
detector components.

Mechanical joints can modify modal damping and introduce
nonlinear hysteretic behavior and microslip
\cite{Mathis2020JointDamping}. Consequently, the
frequencies, mode shapes, and magnetic coupling of
mechanical resonances in the $1$--$20~{\rm kHz}$ search
band should be characterized experimentally.
The measured relative-motion and internal-mode transfer
functions must ultimately be included in the
vibration-induced magnetic-flux noise budget.

\subsection{Persistent-mode decay and field stabilization}
\label{subsec_persistent_decay}

The baseline high-field option is a superconducting magnet operated in persistent-current mode. Persistent operation disconnects the magnet from the power supply and thereby eliminates direct power-supply current noise, but the closed superconducting loop can still exhibit slow current decay due to residual losses. For a constant effective loop inductance $L_m$ and resistance $R_{\rm eff}$, the contribution from persistent-current decay is approximated by
\begin{equation}
\frac{\delta B_{\rm dec}}{B}
\simeq
-\frac{t}{\tau_p}
\quad (t\ll\tau_p),
\qquad
\tau_p=\frac{L_m}{R_{\rm eff}}.
\label{eq_persistent_current_decay}
\end{equation}
This expression describes the resistive decay of the persistent current and does not include other sources of field drift, such as screening-current relaxation or thermal variations.

Long-term persistent-mode operation of superconducting MRI magnets demonstrates that extremely small effective joint resistances and sub-ppm-per-hour drift are achievable. A $0.5~{\rm T}$ whole-body MRI magnet operated in persistent-current mode for 25 years showed a reported field stability of $0.0006~{\rm ppm/h}$ and a total superconducting-joint resistance of $4\times10^{-12}~\Omega$~\cite{Yamamoto2014tasc}. More generally, temporal stability in MRI and NMR superconducting magnets is influenced by persistent-joint resistance, screening-current relaxation in HTS conductors, thermal variations, power-supply ripple in driven-mode magnets, and active field-frequency stabilization where implemented~\cite{Takeda2022sst}. Recent single-ion Penning-trap measurements with a cryogen-free $7~{\rm T}$ superconducting magnet reported a long-term fractional magnetic-field drift of $(-3.07\pm0.32)\times10^{-9}~{\rm h}^{-1}$~\cite{Schupeta2026Cyclotron}. These measurements provide benchmarks for low-frequency field stability but do not establish the magnetic-noise spectrum in the kHz detection band.

For the present gravitational-wave detector, active magnetic-field stabilization must be designed to preserve the gravitational-wave signal transfer function. A feedback loop that senses and suppresses the same flux perturbation used for gravitational-wave detection may also attenuate the signal. In a simple negative-feedback configuration where the gravitational-wave signal and magnetic disturbance enter the same sensed channel, the residual perturbation can be written schematically as
\begin{equation}
\delta\Phi_{\rm res}(\omega)
=
\frac{\delta\Phi_{\rm in}(\omega)}
{1+G(\omega)},
\label{eq_feedback_signal_suppression}
\end{equation}
where $G(\omega)$ is the open-loop gain. In this configuration, $|G(\omega_{\rm GW})|\ll1$ is a sufficient condition for avoiding significant signal attenuation. However, this condition is not universal: a stabilization system using an independent magnetic reference sensor may suppress field noise without applying the same transfer function to the gravitational-wave-induced flux. The relevant requirement is therefore that the calibrated gravitational-wave response be preserved throughout the $1$--$20~{\rm kHz}$ search band. Low-frequency field stabilization may be implemented during data taking if this condition is satisfied.

The raw persistent-mode decay does not need to satisfy the strain-noise {$S_{hh}^{1/2}$} requirement of Eq.~\eqref{eq_db_over_b_hmin_requirement}. Persistent-current decay is predominantly a quasi-DC or very-low-frequency effect, whereas the gravitational-wave search is performed in a narrow kHz band. The relevant requirement is that residual magnetic fluctuations coupled to the pickup remain below the equivalent input-referred strain-noise target in the analysis band, accounting for the antenna response and magnetic-flux coupling. Slow baseline drift can be characterized and removed using time-domain detrending and appropriate frequency-domain filtering, provided that the gravitational-wave signal is preserved. Phase-sensitive demodulation may be used for calibration or for searches with a well-defined signal phase model. A simultaneous reference pickup or dummy channel can also help identify and subtract common-mode magnetic backgrounds, provided that its gravitational-wave response is negligible or explicitly accounted for.

Persistent-current decay can also be compensated by active control within a superconducting circuit. Brouwer et al.~\cite{Brouwer2022sst} demonstrated that a variable superconducting inductance placed in a persistent-current loop can redistribute magnetic flux between the variable inductor and the main magnet, allowing the persistent current to be adjusted or stabilized without an external current supply. Such a technique is not required for the baseline sensitivity estimate, but it provides a possible route to compensate long-term current decay or tune the magnet field while maintaining persistent operation. If implemented during data taking, its sensing and actuation transfer functions must be characterized to ensure that magnetic-noise suppression does not introduce excess noise or attenuate the gravitational-wave response in the search band.

\subsection{Permanent magnets}
\label{subsec_permanent_magnet}

Permanent magnets provide an attractive alternative for lower-field prototypes. They require no power supply, have no persistent-current decay, cannot quench, and can be compact and mechanically rigid. Cryogenic Halbach-type assemblies have demonstrated fields up to the tesla scale with drift below a few ppb per hour in dilution-refrigerator operation~\cite{Adambukulam2020permanent}. These properties make permanent magnets useful for engineering tests of the LC resonator, vibration coupling, reference subtraction, and lock-in analysis.

Their limitations are equally important. The attainable field is typically well below the $9$--$14~{\rm T}$ range assumed for the final high-sensitivity configuration, the available field volume is small, the gradient can be much larger than in a shimmed superconducting solenoid, and the field is sensitive to temperature-dependent magnetization and mechanical motion. Since the gravitational-wave flux drive scales as $B$, replacing a $14~{\rm T}$ superconducting magnet by a $1.5~{\rm T}$ permanent magnet reduces the signal coefficient by approximately a factor of nine. A permanent magnet is therefore a valuable prototype platform, but it is not the preferred final magnet for the highest-sensitivity configuration unless its lower field is compensated by a substantially lower vibration and magnetic-noise budget.

\section{Conclusions}
\label{sec_conclusions}

We have presented a compact superconducting LC-resonator concept for gravitational-wave detection in the kilohertz band, based on the proposed conversion of gravitational-wave-induced electromagnetic perturbations into magnetic flux in a superconducting pickup circuit immersed in a multi-tesla magnetic field. Building on the coupled-mode SQUID readout developed for resonant-mass detectors, we formulated a three-loop detector--readout--SQUID model and derived the corresponding signal transfer function and input-referred strain-noise spectrum, including detector thermal noise, readout thermal noise, SQUID backaction, and measurement imprecision. Measurements of a prototype LC resonator demonstrated an intrinsic quality factor of $Q\simeq5\times10^5$ near $1.5~{\rm kHz}$, while $Q_d=Q_A=10^7$ remains a future development target for the projected detector configuration. Under the assumed gravitational-wave-to-flux coupling, high-field geometry, $Q_d=Q_A=10^7$, and idealized quantum-limited SQUID noise model, the optimized circuits yield projected strain-noise amplitude spectral densities of order $10^{-21}~{\rm Hz}^{-1/2}$ over approximately $100~{\rm Hz}$ bands at selected frequencies between $1$ and $20~{\rm kHz}$. Further cooling alone provides only modest improvement once SQUID-added noise dominates.

Additionally, the magnet and mechanical requirements are essential to the realization of this concept. A high static magnetic field enhances the assumed gravitational-wave flux response, but its associated magnetic noise must also be controlled. Field gradients and relative motion between the magnet and pickup circuit, together with deformations of the pickup and surrounding structures, can generate spurious magnetic-flux signals. These effects require magnetic-field mapping, mechanical characterization, vibration isolation, and tests for nonlinear frequency upconversion. Persistent-current operation can reduce direct power-supply noise, while stabilization and reference-channel subtraction must be designed to preserve the gravitational-wave signal transfer function. In particular, low-frequency magnetic-field drift and magnetic noise in the kilohertz detection band must be evaluated separately.

In the future, greater sensitivity could be pursued through a larger active high-field volume, improved resonator performance, and lower-added-noise readout techniques, including phase-sensitive or backaction-evading measurement schemes \cite{Caves1982,Clerk2010rmp,Kuenstner2025RQU}. The predicted sensitivity remains conditional on experimental validation of the gravitational-wave flux response, the high-field resonator quality factor, and the magnetic and mechanical noise budget. Subject to these developments, a compact, tunable superconducting LC detector could complement interferometric observations in the kilohertz regime and motivate targeted searches for signals associated with neutron-star post-merger remnants, subsolar-mass compact-object mergers, stellar collapse, and ultralight boson clouds around rotating black holes.

\begin{acknowledgments}
We are grateful to E. Berto, A. Benato, and M. Rebeschini
for the mechanical work; F. Calaon and M. Tessaro for
help with the electronics and cryogenics.
We also thank Kent Irwin for valuable discussions on the
SQUID readout circuit.
{We gratefully acknowledge} the Cryogenic Service of the
Laboratori Nazionali di Legnaro (LNL) for providing us with liquid helium on demand.
This work is supported by Commissione Scientifica Nazionale 2 (CSN2) of the Istituto Nazionale di Fisica Nucleare (INFN) {for the QUaerere AXion (QUAX) experiment}.
OpenAI ChatGPT (GPT-5.6 Sol) was used to assist with manuscript drafting and language editing. All AI-assisted text was reviewed, verified, and revised by the authors, who take full responsibility for the content.
\end{acknowledgments}

\section*{Data Availability}
The data that support the findings of this article are not publicly available upon publication because it is not technically feasible and/or the cost of preparing, depositing, and hosting the data would be prohibitive within the terms of this research project. The data are available from the authors upon reasonable request.

\bibliography{references}

\end{document}